\documentclass[12pt]{article}
\usepackage{amssymb}
\usepackage{epsfig}
\usepackage{amssymb,amsmath}
\usepackage{epstopdf}
\hoffset= - 0.8in         

\catcode`\@=11
\def\marginnote#1{}

\newcount\hour
\newcount\minute
\newtoks\amorpm
\hour=\time\divide\hour by60
\minute=\time{\multiply\hour by60 \global\advance\minute by-\hour}
\edef\standardtime{{\ifnum\hour<12 \global\amorpm={am}%
        \else\global\amorpm={pm}\advance\hour by-12 \fi
        \ifnum\hour=0 \hour=12 \fi
        \number\hour:\ifnum\minute<10 0\fi\number\minute\the\amorpm}}
\edef\militarytime{\number\hour:\ifnum\minute<10 0\fi\number\minute}

\def\draftlabel#1{{\@bsphack\if@filesw {\let\thepage\relax
   \xdef\@gtempa{\write\@auxout{\string
      \newlabel{#1}{{\@currentlabel}{\thepage}}}}}\@gtempa
   \if@nobreak \ifvmode\nobreak\fi\fi\fi\@esphack}
        \gdef\@eqnlabel{#1}}
\def\@eqnlabel{}
\def\@vacuum{}
\def\draftmarginnote#1{\marginpar{\raggedright\scriptsize\tt#1}}

\def\draft{\oddsidemargin -.5truein
        \def\@oddfoot{\sl preliminary draft \hfil
        \rm\thepage\hfil\sl\today\quad\militarytime}
        \let\@evenfoot\@oddfoot \overfullrule 3pt
        \let\label=\draftlabel
        \let\marginnote=\draftmarginnote
   \def\@eqnnum{(\theequation)\rlap{\kern\marginparsep\tt\@eqnlabel}%
\global\let\@eqnlabel\@vacuum}  }

\def\d{\partial}

\def\bea{\begin{eqnarray}}
\def\eea{\end{eqnarray}}

\def\beq{\begin{equation}}
\def\eeq{\end{equation}}
\def\ba{\beq\new\begin{array}{c}}
\def\ea{\end{array}\eeq}
\def\be{\ba}
\def\ee{\ea}
\def\stackreb#1#2{\mathrel{\mathop{#2}\limits_{#1}}}
\def\Tr{{\rm Tr}}
\def\Ad{{\rm Ad}}
\def\sign{{\rm sign}}

\makeatletter
\newdimen\normalarrayskip              
\newdimen\minarrayskip                 
\normalarrayskip\baselineskip
\minarrayskip\jot
\newif\ifold             \oldtrue            \def\new{\oldfalse}
\def\arraymode{\ifold\relax\else\displaystyle\fi} 
\def\eqnumphantom{\phantom{(\theequation)}}     
\def\@arrayskip{\ifold\baselineskip\z@\lineskip\z@
     \else
     \baselineskip\minarrayskip\lineskip2\minarrayskip\fi}
\def\@arrayclassz{\ifcase \@lastchclass \@acolampacol \or
\@ampacol \or \or \or \@addamp \or
   \@acolampacol \or \@firstampfalse \@acol \fi
\edef\@preamble{\@preamble
  \ifcase \@chnum
     \hfil$\relax\arraymode\@sharp$\hfil
     \or $\relax\arraymode\@sharp$\hfil
     \or \hfil$\relax\arraymode\@sharp$\fi}}
\def\@array[#1]#2{\setbox\@arstrutbox=\hbox{\vrule
     height\arraystretch \ht\strutbox
     depth\arraystretch \dp\strutbox
     width\z@}\@mkpream{#2}\edef\@preamble{\halign
\noexpand\@halignto
\bgroup \tabskip\z@ \@arstrut \@preamble \tabskip\z@ \cr}%
\let\@startpbox\@@startpbox \let\@endpbox\@@endpbox
  \if #1t\vtop \else \if#1b\vbox \else \vcenter \fi\fi
  \bgroup \let\par\relax
  \let\@sharp##\let\protect\relax
  \@arrayskip\@preamble}
\def\eqnarray{\stepcounter{equation}%
              \let\@currentlabel=\theequation
              \global\@eqnswtrue
              \global\@eqcnt\z@
              \tabskip\@centering
              \let\\=\@eqncr
 \halign to \displaywidth\bgroup
    \eqnumphantom\@eqnsel\hskip\@centering
    $\displaystyle \tabskip\z@ {##}$%
    \global\@eqcnt\@ne \hskip 2\arraycolsep
         $\displaystyle\arraymode{##}$\hfil
    \global\@eqcnt\tw@ \hskip 2\arraycolsep
         $\displaystyle\tabskip\z@{##}$\hfil
         \tabskip\@centering
    &{##}\tabskip\z@\cr}
\begingroup\ifx\undefined\newsymbol \else\def\input#1 {\endgroup}\fi
\newfont{\hr}{msbm10}
\newfont{\ams}{msam10}

\font\numbers=cmss12
\font\upright=cmu10 scaled\magstep1
\def\stroke{\vrule height8pt width0.4pt depth-0.1pt}
\def\topfleck{\vrule height8pt width0.5pt depth-5.9pt}
\def\botfleck{\vrule height2pt width0.5pt depth0.1pt}
\def\Zmath{\vcenter{\hbox{\numbers\rlap{\rlap{Z}\kern 0.8pt\topfleck}\kern
2.2pt
                   \rlap Z\kern 6pt\botfleck\kern 1pt}}}
\def\Qmath{\vcenter{\hbox{\upright\rlap{\rlap{Q}\kern
                   3.8pt\stroke}\phantom{Q}}}}
\def\Nmath{\vcenter{\hbox{\upright\rlap{I}\kern 1.7pt N}}}
\def\Cmath{\vcenter{\hbox{\upright\rlap{\rlap{C}\kern
                   3.8pt\stroke}\phantom{C}}}}
\def\Rmath{\vcenter{\hbox{\upright\rlap{I}\kern 1.7pt R}}}
\def\Z{\ifmmode\Zmath\else$\Zmath$\fi}
\def\Q{\ifmmode\Qmath\else$\Qmath$\fi}
\def\N{\ifmmode\Nmath\else$\Nmath$\fi}
\def\C{\ifmmode\Cmath\else$\Cmath$\fi}
\def\R{\ifmmode\Rmath\else$\Rmath$\fi}

\def\stackreb#1#2{\mathrel{\mathop{#2}\limits_{#1}}}
\def\Tr{{\rm Tr}}

\def\Bf#1{\mbox{\boldmath $#1$}}

\def\bmu{{\Bf\mu}}

\def\d{\partial}

\def\rank{{\rm rank}}
\def\diag{{\rm diag}}
\def\half{{\textstyle{1\over2}}}

\def\2{{1\over 2}}

\def\beq{\begin{equation}}
\def\eeq{\end{equation}}
\def\ba{\beq\new\begin{array}{c}}
\def\ea{\end{array}\eeq}
\def\be{\ba}
\def\ee{\ea}
\def\stackreb#1#2{\mathrel{\mathop{#2}\limits_{#1}}}

\newcommand{\rf}[1]{(\ref{#1})}

\begin{document}


\begin{flushright}
FIAN/TD-03/14\\
ITEP/TH-09/14
\end{flushright}
\vspace{1.0 cm}
\renewcommand{\thefootnote}{\fnsymbol{footnote}}
\begin{center}
\baselineskip20pt
{\bf \LARGE On Lie Groups and Toda Lattices}
\end{center}
\bigskip
\begin{center}
\baselineskip12pt
{\large O.~Kruglinskaya and A.~Marshakov}\\
\bigskip
{\em Theory Department, Lebedev Physics Institute,\\
Institute for Theoretical and Experimental Physics,\\
Department of Mathematics and Laboratory of\\
Mathematical Physics, NRU HSE,
Moscow, Russia}\\
\end{center}
\bigskip\medskip

\begin{center}
{\large\bf Abstract} \vspace*{.2cm}
\end{center}

\begin{quotation}
\noindent
We extend the construction of the relativistic Toda chains as integrable systems on the Poisson submanifolds in Lie groups beyond the case of A-series. For the simply-laced case
this is just a direct generalization of the well-known relativistic Toda chains, and we construct explicitly the set of the Ad-invariant integrals
of motion on symplectic leaves, which can be described by the Poisson quivers being just the blown up Dynkin diagrams. We also demonstrate how to get the set of ``minimal'' integrals of motion, using the
co-multiplication rules for the corresponding Lie algebras.
In the non simply-laced case the corresponding Bogoyavlensky-Coxeter-Toda systems are constructed
using the Fock-Goncharov folding of the corresponding Poisson submanifolds. We discuss also how this procedure can be extended for the affine case beyond A-series, and consider explicitly an example from the affine D-series.
\end{quotation}

\newpage

\renewcommand{\thefootnote}{\arabic{footnote}}
\setcounter{section}{0}
\setcounter{footnote}{0}
\setcounter{equation}0
\section{Introduction}

The relation to Lie groups is one of the basic stones of the theory of integrable systems, see e.g.  \cite{Sembook}. It comes out in a particularly straightforward way \cite{FMold}, when an integrable model can be directly constructed on a Poisson submanifold in Lie group. Integrability in this case
just follows from existence of the Ad-invariant functions on group
manifold, which are in the involution w.r.t. to the r-matrix Poisson bracket, consistent with the group multiplication law.

This relation has been studied in detail \cite{AMJGP,FM}
and extended to the affine Lie groups of the $A_{N-1}$ series, where it gives
rise to a nontrivial wide class of integrable models, alternatively
discovered in \cite{GK} as arising from the dimer models on bipartite graphs on a torus.
This has been done effectively due to ability to use the language of cluster varieties \cite{cv}
or cluster algebras, as they are called more often. In cluster coordinates the Poisson bracket becomes
logarithmically constant and can be encoded by the Poisson graph (or Poisson quiver) with cluster
coordinates attached to its vertices, while the integrals of motion turn to be the cluster functions,
and can be computed explicitly, using the so called Lax map.

There is certainly not a unique way to introduce cluster structure on the Poisson submanifolds
in Lie groups, which can be also viewed as double Bruhat cells \cite{bruh,Resh}. A particular choice
of a system of cluster coordinates is mostly a question of personal taste or convenience: for
example slightly different, but related languages to what we are going to use below can be found,
for example, in \cite{GSW,GSWbook,HW}, and in \cite{pent,SolKhe}, where they are adjusted
to study the discrete flows in such systems, mostly known from which is the pentagram map. We
are going to describe the Poisson submanifolds as cluster $\mathcal{X}$-varieties, following the amalgamation procedure proposed in \cite{DrinFG}, which turns to be the most transparent for the formulation of relativistic Toda systems
on the symplectic leaves in $G/\mbox{Ad}H$.

The main aim of this paper is show, that this language can be directly extended
to the other series of Lie groups (while some other choices of coordinates can be more useful to
study, for example, different cluster structure on $SL(N)$, see \cite{GSW}).
We first show, that the construction of \cite{AMJGP,FM} can be applied almost without modifications
to all simple Lie groups of the ADE-series, resulting in relativistic generalizations \cite{Ruj,Suris}
of the generalized Bogoyavlensky-Coxeter-Toda lattices. Integrability still follows from the simple
counting argument, which ensures existence of exactly $\mbox{rank}G$ independent $\mbox{Ad}$-invariant integrals of motion on symplectic list of dimension $2\cdot\mbox{rank}G$. The only extra question one
can ask here is how to get the set of ``minimal'' integrals, and we demonstrate below, that they are still given by traces in fundamental representations, while their explicit form can be extracted from the coefficients of characteristic polynomial of the matrix in vector representation, using the co-multiplication rules for a group $G$.

We also demonstrate, that the integrable Toda systems on the appropriate symplectic leaves of non simply-laced
simple groups can be obtained from the previous class just by folding of the corresponding cluster variety. This is a straightforward procedure, described for the $\mathcal{X}$-submanifolds in \cite{DrinFG} just
as gluing of the corresponding Poisson graphs. We will show, that being applied to the Toda symplectic leaves of the simply-laced groups it gives rise immediately to the generalized Toda systems of the
non-ADE series, preserving all elements of the construction - and the simplest way to get explicitly
the expressions for the integrals of motion in cluster coordinates is still to use the Lax map.

Finally, we conjecture, that proposed scheme can be extended also to the affine case. The detailed study
of this class goes far beyond the scope of this paper. However, we shall derive the shape of the spectral curves for the affine $\hat{D}$-Toda systems, and show that already for this simplest subclass of the Poisson submanifolds in affine groups of $\hat{D}$-type, the construction of \cite{FM} requires some modification.

\setcounter{equation}0
\section{Integrable system
\label{ss:reltoda}}

It is easy \cite{FMold} to see directly from the $r$-matrix Poisson bracket
\be\label{rbra}
\{ g \stackreb{,}{\otimes}g \} = -\half\ [r,g\otimes g]
\ee
that any two $\Ad$-invariant functions on a simple Lie group $G$ do Poisson-commute with each other. For any $r$-matrix $r\in \mathfrak{g}\otimes \mathfrak{g}$,
where $\mathfrak{g} = {\rm Lie}(G)$, e.g. antisymmetric - given by
\be
\label{ref}
r =  \sum_{\alpha\in\Delta _+} d_\alpha e_\alpha \wedge e_{\bar\alpha} =
\sum _{\alpha\in\Delta _+}d_\alpha\left( e_{\alpha}\otimes e_{\bar\alpha} -
e_{\bar\alpha}\otimes e_{\alpha}\right)
\ee
(where $d_\alpha = \half(\alpha,\alpha)$, the sum is taken over the set of all positive roots $\Delta_+$ and by $e_{\bar\alpha} = f_\alpha$ we denoted the corresponding negative roots, see Appendix~\ref{ap:promunu} for more details about accepted notations), the Poisson bracket on $G$ is defined by
\be
\label{gpb}
\{\mathcal{H}_1,\mathcal{H}_2\}=-\half\sum_{\alpha\in\Delta _+}d_\alpha\left(L_{e_\alpha}\mathcal{H}_1L_{e_{\bar\alpha}}\mathcal{H}_2-
R_{e_\alpha}\mathcal{H}_1R_{e_{\bar\alpha}}\mathcal{H}_2\right)
\ee
where for any $v\in \mathfrak{g}$ we denote by $L_v$ ($R_v$) the corresponding left (right) vector field.
Any $\Ad$-invariant function $\mathcal{H}$ satisfies $L_v \mathcal{H} = -R_v \mathcal{H}$ and thus the bracket \rf{gpb} of two such functions vanishes.
The bracket \rf{gpb} obviously vanishes even if the functions are defined not on the whole $G$, but on any Poisson $\Ad$-invariant subvariety of $G$.

On a simple group there exists exactly $\rank\ G$ independent $\Ad$-invariant functions: a possible basis of these functions is the set $\{\mathcal{H}_i\}$, where $i \in \Pi$  (the set of positive simple roots $\Pi\subset\Delta_+$)
\be
\label{trfund}
\mathcal{H}_i = \Tr\ \pi_{\mu_i}(g)
\ee
and $\pi_{\mu_i}$ be the $i$-th fundamental representation of $G$ with the highest weight $\langle \alpha_i,\mu_j\rangle=\delta_{ij}$ dual to $\alpha_i$, $i\in\Pi$. These functions then define a completely integrable system on a symplectic leaf of dimension $2\cdot \rank\ G$, which for $G=SL(N)$ turns to be
the relativistic Toda chain \cite{Ruj,Suris}, with
the canonical Hamiltonian
\be
\label{HslN}
{\cal H} = \Tr\left(g+g^{-1}\right) = \sum_{i=1}^N\left(\exp(p_i)+\exp(-p_i)\right)\sqrt{1+\exp(q_i-q_{i+1})}\sqrt{1+\exp(q_{i-1}-q_i)}
\ee
while
more known canonical (non-relativistic) Toda system is recovered in the limit from the group $G=SL(N)$ to its Lie algebra (see, for example \cite{AMJGP}).

It has been also shown \cite{AMJGP,FM}, that the most convenient way to construct the integrals of motion comes out of using the cluster $\mathcal{X}$-coordinates  on the double Bruhat cells in Lie groups \cite{bruh} and on the Poisson submanifolds in $G/\mbox{Ad}H$. The Poisson structure in these cluster coordinates can be then described in terms of the Poisson quiver, which can be obtained by the amalgamation procedure of Fock and Goncharov \cite{DrinFG}.

\section{Simply-laced case
\label{ss:ade}}

\begin{figure}[t]
\center{\includegraphics[width=14.5cm]{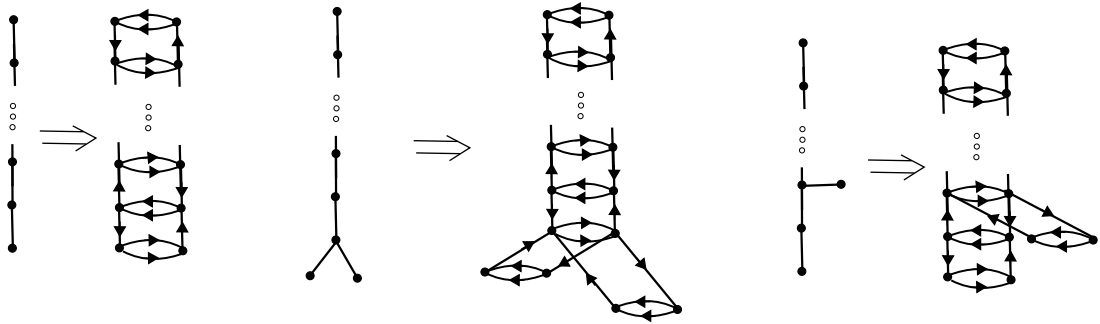}}
\caption{\sl Poisson graphs for the ADE-series as blowups of the corresponding Dynkin diagrams.}
\label{fi:blowup}
\end{figure}

In the simply-laced case the Poisson quivers of the Toda symplectic leaves in $G/\mbox{Ad}H$ can be drawn just by ''blowup'' of the corresponding Dynkin diagrams, see fig.~\ref{fi:blowup}. On the left picture for the $A$-series such graph is just a ladder, constructed
from the Dynkin chain, and the $D$- and $E$-series differ only by existence of a junction on the
corresponding Dynkin diagram - and, correspondingly, on the ladder. The coordinates on the corresponding leaf are associated with the vertices of the graph, and the Poisson structure is defined in this coordinates by
\be
\label{clbra}
\{ z_I,z_J\} = \varepsilon_{IJ}z_Iz_J,\ \ \ \ I,J=1,\ldots,\# V
\ee
where $\varepsilon_{IJ}$ stays for the exchange matrix
\be
\label{ince}
\varepsilon_{IJ} = \# {\rm arrows}\ (I\rightarrow J) - \# {\rm arrows}\ (J\rightarrow I)
\ee
of the Poisson quiver. In principle, each such graph defines a Poisson manifold $\mathbb{C}^{\# V}$
with logarithmically constant bracket \rf{clbra}.

It is seen from fig.~\ref{fi:blowup}, that such graphs for the Toda symplectic leaves of dimension $\# V = 2\cdot\mbox{rank}G$ define the Poisson bracket, which can be conveniently written as (see \cite{AMJGP})
\be
\label{pbN}
\{ y_i,x_j\}_{\mathfrak{g}} = C_{ij}^{\mathfrak{g}}y_ix_j,\ \ \ i,j=1,\ldots,\mbox{rank}G=N-1
\\
\{ x_i,x_j\}_{\mathfrak{g}} = 0,\ \ \ \{ y_i,y_j\}_{\mathfrak{g}} = 0
\ee
for $\mathfrak{g}=A_{N-1}$, $\mathfrak{g}=D_{N-1}$ or $\mathfrak{g}=E_{N-1}$, where $C_{ij}^{\mathfrak{g}}$ is the corresponding Cartan matrix (e.g. for $A_{N-1}$ it can be explicitly written in the form of \rf{CslN}).  Natural splitting of all variables $\{z_I\}$ from \rf{clbra} as $\{z_I\} = \{ \mathbf{x},\mathbf{y}\}$ is useful only on symplectic leaves, and formulas \rf{pbN} show, that they are easily expressed through the Darboux coordinates, as we will see below.

Moreover, it has been shown in \cite{DrinFG}, that the Poisson bracket \rf{pbN} is just a restricted r-matrix bracket \rf{rbra} on Lie group, if the group element is parameterized in the form
\be
\label{gADE}
g({\bf x},{\bf y})
\simeq \prod_{i=1}^{N-1}E_iH_i(x_i)E_{\bar i}H_i(y_i)
\ee
where (see Appendix~\ref{ap:promunu})
\be
E_i=\exp (e_i),\ \ \ \ E_{\bar i}=\exp (e_{\bar i}),\ \ \ \ H_i(z) = \exp(h^i\log z)
\ee
are the exponentiated Chevalley generators, taken in the vector representation of $G$.
It is easy to see, that the order of
factors in \rf{gADE} is inessential: many of them are equivalent up to a cyclic permutation (which remains intact all Ad-invariant functions), while the rest correspond to the same Poisson submanifolds, up to a cluster transformation (a sequence of mutations \cite{cv}). It has been shown in \cite{FMold}, that matrices in \rf{rbra} play the role of the Lax operators for the corresponding integrable systems, so
that the formulas \rf{gADE} give their convenient parametrization in cluster coordinates \cite{AMJGP,FM}. We are going to exploit this parametrization below in order to construct explicit formulas for the integrals of motion and study their properties.

In the $A_{N-1}$ case the characteristic polynomial for the Lax operator \rf{gADE}
\be
\label{RslN}
\det \left(\mu+g({\bf x},{\bf y})\right) =
\sum_{j=0}^N\mu^jR_j({\bf x},{\bf y})
\ee
generates $\rank\ SL(N) = N-1$ nontrivial ($R_0=R_N=1$ in the accepted normalization,
and $R_1 = \Tr_V g^{-1}=\mathcal{H}_{N-1}$, $R_{N-1}= \Tr_V g = \mathcal{H}_1$, see \cite{AMJGP}) Poisson-commuting (w.r.t. the bracket \rf{pbN}) integrals of motion $\{ \mathcal{H}_i,\mathcal{H}_j\} = 0$, $i,j=1,\ldots,N-1$, which read
\be
\label{Rhamsl}
\mathcal{H}_j({\bf x},{\bf y}) = \prod_k \left(x_ky_k\right)^{-C^{-1}_{jk}}\cdot Z_j({\bf x},{\bf y})
= R_{N-j}({\bf x},{\bf y})
\ee
where, for the $A$-series
\be
\label{dimer}
Z_j({\bf x},{\bf y}) = \sum_{0\le m_i\le \max(i,N-1-i)}^{m_j\ge m_{j\pm 1}\ge m_{j\pm 2}\ge\ldots}\
\sum_{m_i-1\le n_i\le m_i}\
\prod_i y_i^{m_i}x_i^{n_i}
\ee
are already polynomials with all unit coefficients. The nice structure of these formulas (\rf{Rhamsl} are
cluster functions and the Laurent polynomials in fractional powers of the cluster variables) has elegant
interpretation in terms of the quiver representation theory \cite{HWN}, however - to our knowledge - it has no generalization yet to the analogous formulas for the other groups, which we obtain below.

The coefficients of characteristic polynomial \rf{RslN} in the $SL(N)$ case
directly give the set of ``minimal'' integrals of motion, since $R_{N-j}({\bf x},{\bf y}) = \Tr_{\mu_j} g({\bf x},{\bf y})$ with $j=1,\ldots,N-1$ just correspond to the traces in all fundamental representations, labeled by the
fundamental weights $(\alpha_i,\mu_j)=\delta_{ij}$, $i,j\in\Pi$.

The true Darboux coordinates are related to the cluster variables $\{z_I\} = \{ \mathbf{x},\mathbf{y}\}$ by
\be
\label{darbun}
x_i = \exp(-(\alpha_i\cdot q)),
\ \ \ \
y_i = \exp((\alpha_i\cdot P)+(\alpha_i\cdot q)),\ \ \ i=1,\ldots,N-1
\ee
with the ``long momentum''
\be
\label{Pp}
P = p - {1\over 2}\sum_{k=1}^{N-1}\alpha_k \log\left(1+\exp(\alpha_k\cdot q)\right) =
p + {\d\over\d q}\left( {1\over 2}\sum_{k=1}^{N-1}{\rm Li}_2\left(-\exp(\alpha_k\cdot q)\right)\right)
\ee
where $(N-1)$-vectors $q$, $p$, denote the canonical coordinate and momentum in the center of mass frame.
Substituting \rf{darbun} into the expression
$\mathcal{H}= \mathcal{H}_1+\mathcal{H}_{N-1}=R_1 + R_{N-1}$ one gets the
canonical relativistic Toda Hamiltonian \rf{HslN}, where one has to drop
off the square roots with $\alpha_0$ and $\alpha_N$,
i.e. just to replace $\sqrt{1+\exp(q_0-q_1)}$ and $\sqrt{1+\exp(q_N-q_{N+1})}$ by unities.

For the other simply-laced groups (of $D$- and $E$- series) the Lax operator \rf{gADE} gives rise instead of \rf{RslN} an equation
\be
\label{RDE}
\det \left(\mu+g({\bf x},{\bf y})\right) =
\sum_{j=0}^{\dim V}\mu^jR_j({\bf x},{\bf y})
\ee
for a polynomials of higher power in $\mu$, since $\dim V > \rank\ G +1$ in contrast to the $SL(N)$
groups. All coefficients of this equation are still the integrals of motion, being in the involution
$\{ R_i,R_j\}_{\mathfrak{g}} =0$ w.r.t. the Poisson bracket \rf{pbN}
for all $i,j=1,\ldots,\dim V-1$ ($R_0=R_{\dim V}=1$), just because they are Ad-invariant functions, and
the bracket \rf{pbN} is equivalent to \rf{rbra}.

However, they are now neither all independent, nor minimal - except for the few ones, as we see below.
In order
to find the full set of minimal integrals of motion, one should use the relations between the wedge
powers $\wedge^j V$ of the vector representation, corresponding to the coefficients of the
characteristic polynomial \rf{RDE} with the fundamental representations of $G$, this can be done, for
example, with the help of \cite{vinb,prog}: it is more instructive to demonstrate this on particular examples, and we take one at a time for $D$ and $E$ cases.

\subsection{$D_4$ Toda and triality
\label{ss:D_4}}

This is perhaps the simplest case out of common $SL(N)$ Toda systems, and it is also
rather interesting to observe the famous triality, coming from $\mathbb{Z}_3$-symmetry of the Dynkin diagram. It predicts existence of three ``similar'' minimal integrals of motion, which comes in
conflict with naive analysis of the characteristic polynomial equation \rf{RDE}.

\begin{figure}[t]
\center{\includegraphics[width=14.5cm]{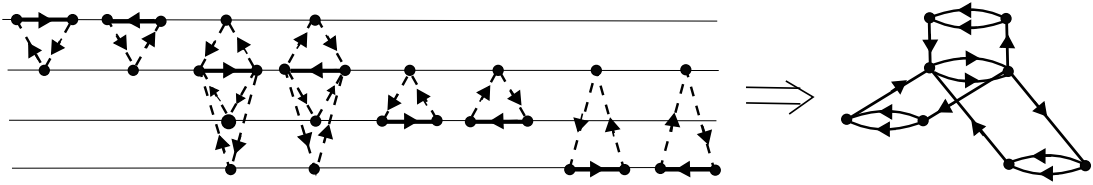}}
\caption{\sl Construction of a Poisson quiver for the Toda symplectic leaf in $D_4=SO(8)$.}
\label{fi:todaD4}
\end{figure}

First, we present at fig.~\ref{fi:todaD4} the amalgamation construction of \cite{DrinFG} for the $SO(8)$-Toda symplectic leaf. As was announced it results in blow-up of a ``sicilian'' Dynkin diagram
of $D_4$, and the corresponding Poisson graph is glued from eight constituents - cluster seeds,
each of which corresponds
to positive ($i\in\Pi$) or negative ($i\in\bar{\Pi}$) simple root (see notations in Appendix~\ref{ap:promunu}). Literally picture from fig.~\ref{fi:todaD4} describes a
symplectic leaf, given by the Coxeter-Toda word $u=1\bar{1}2\bar{2}3\bar{3}4\bar{4}$, where bars denote negative roots, and numbers label the levels from up to down. It means, in particular, that triality should permute the variables, corresponding to the first, third and fourth roots.

For the Lax operator \rf{gADE} one considers therefore the product
\be
\label{prod4}
g({\bf x},{\bf y})
= \underbrace{E_1X_1E_{\bar 1}Y_1}_{g_1}\cdot
\underbrace{E_2X_2E_{\bar 2}Y_2}_{g_2}\cdot\underbrace{E_3X_3E_{\bar 3}Y_3}_{g_3}\cdot\underbrace{E_4X_4E_{\bar 4}Y_4}_{g_4}
\\
Y_i = H_i(y_i),\ \ \
X_i = H_i(x_i),\ \ \ i=1,\ldots,4
\ee
of the matrices in 8-dimensional vector representation of $SO(8)$, where one can take
\be
\label{ED4}
E_1 = \mathbb{I}+e_{12}-e_{78}, \; \;
E_2 = \mathbb{I}+e_{23}-e_{67},
\\
E_3= \mathbb{I}+e_{34}-e_{56}, \; \;
E_4 = \mathbb{I}+e_{35}-e_{46}
\ee
(and $E_{\bar i}=E_i^{\rm tr}$),
with $\mathbb{I}$ being the identity matrix, while $e_{ij}$ denotes the matrix
$\|e_{ij} \|_{kl} = \delta_{ik}\delta_{jl}$
with the only unit element on intersection of $i$-th row with $j$-th column. The parametrization of the Lax operator by cluster coordinates is induced through the Cartan elements, which we take as usual as $H_i(z) = \exp(h^i\log z)$ (see notations in Appendix~\ref{ap:promunu}), $i=1,\ldots,4$, i.e.
\be
H_1(z)=\diag \{ z,1,1,1,1,1,1,1/z \}, \; H_2(z)=\diag \{ z,z,1,1,1,1,1/z,1/z \},
\\
H_3(z)=\diag \{ \sqrt{z}, \sqrt{z}, \sqrt{z}, 1/\sqrt{z}, \sqrt{z}, 1/\sqrt{z}, 1/\sqrt{z}, 1/\sqrt{z} \},
\\
H_4(z)=\diag \{ \sqrt{z},\sqrt{z},\sqrt{z},\sqrt{z},1/\sqrt{z},1/\sqrt{z},1/\sqrt{z},1/\sqrt{z} \}
\ee
The characteristic polynomial for \rf{prod4} has the form
\be
\label{LaxeqD4}
\det \left(g({\bf x},{\bf y}) + \mu\right) = \mu^8 + \sum_{j=1}^7\mu^jR_j({\bf x},{\bf y}) + 1
\ee
and gives naively seven functions in the involution. However, it is easy to see, that there are only four independent coefficients, and moreover, they can be expressed as
\be
\label{Rr}
R_1 = R_7 = \mathcal{H}_1, \; R_2=R_6 = \mathcal{H}_2,
\\
R_3 = R_5 = \mathcal{H}_3\mathcal{H}_4 - \mathcal{H}_1,
\\
R_4 = \mathcal{H}_3^2+\mathcal{H}_4^2 - 2R_2-2
\ee
through the traces in four fundamental representations $\bmu_k$,
\be
\label{imod4}
\mathcal{H}_k = \Tr_{\bmu_k}g({\bf x},{\bf y}),\ \ \ \ k=1,\ldots,4
\ee
Relations \rf{Rr} correspond exactly to the following co-multiplication rules for $D_4$ (see \cite{vinb,prog}):
\be
\label{comultD4}
V = V(\bmu_1), \;
\wedge^2 V = V(\bmu_2),
\\
\wedge^3 V = V(\bmu_3+\bmu_4) = V(\bmu_3)\otimes V(\bmu_4) - V,
\\
\wedge^4 V = V(2\bmu_3) + V(2\bmu_4) = V(\bmu_3)\otimes V(\bmu_3) + V(\bmu_4)\otimes V(\bmu_4) - 2V(\bmu_2) - 2V_0
\ee
where by $V_0$ and $V$ we have denoted the trivial and $8$-dimensional vector representations correspondingly.

Formulas \rf{Rr} do the last piece of the job: they give the relation between the easily computable coefficients of the characteristic polynomials and the set of minimal integrals of motion.
For \rf{imod4} with $k=1,3,4$ one can write
\be
\label{rfun}
\mathcal{H}_1 = \frac{1}{y_1x_1y_2x_2\sqrt{y_3x_3}\sqrt{y_4x_4}}
(1+y_1+y_1x_1+y_1x_1y_2+ y_1x_1y_2x_2+y_1x_1y_2x_2y_3+
\\
+y_1x_1y_2x_2y_4
+ y_1x_1y_2x_2y_3x_3+y_1x_1y_2x_2y_4x_4+y_1x_1y_2x_2y_3y_4+ y_1x_1y_2x_2y_3x_3y_4+
\\
+y_1x_1y_2x_2y_4x_4y_3+
 y_1x_1y_2x_2y_3x_3y_4x_4+y_1x_1y_2^2x_2y_3x_3y_4x_4
 +y_1x_1y_2^2x_2^2y_3x_3y_4x_4+
 \\
 +y_1^2x_1y_2^2x_2^2y_3x_3y_4x_4+y_1^2x_1^2y_2^2x_2^2y_3x_3y_4x_4 ) =
\\
= \mathfrak{h}(x_1,y_1|x_3,y_3;x_4,y_4|x_2,y_2)
\ee
and
\be
\mathcal{H}_3=\mathfrak{h}(x_3,y_3|x_1,y_1;x_4,y_4|x_2,y_2)
\\
\mathcal{H}_4=\mathfrak{h}(x_4,y_4|x_1,y_1;x_3,y_3|x_2,y_2)
\ee
as predicted by triality (notice, that $\mathfrak{h}(\bullet|a,b;\alpha,\beta|\bullet)=\mathfrak{h}(\bullet|\alpha,\beta;a,b|\bullet)$ is symmetric w.r.t. permutation of the corresponding pairs of variables). The explicit formula \rf{R2D4} for
$\mathcal{H}_2 = R_2 = \Tr_{\bmu_2}g({\bf x},{\bf y})$ can be found in Appendix~\ref{ap:explicit}. Let us point out, that expressions for the integrals of motion still have the form of \rf{Rhamsl}, where the
polynomials have all unit coefficients, which is a distinguishing feature of the simply-laced case.
The canonical Hamiltonian in this case should be $\mathcal{H}=\mathcal{H}_1+\mathcal{H}_3+\mathcal{H}_4$.

\subsection{$E_6$ Toda chain
\label{ss:E6}}

\begin{figure}[t]
\center{\includegraphics[width=5cm]{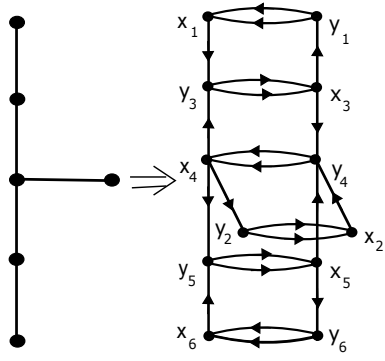}}
\caption{\sl Construction of a Poisson graph for the Toda symplectic leaf in $E_6$.}
\label{fi:todaE6}
\end{figure}

Let us now turn to the simplest example of exceptional series.
In the $E_6$ case instead of \rf{ED4} one can use the matrices in the 27-dimensional vector representations, for example
\be
\label{EE6}
E_1 = \mathbb{I}+e_{1,2}+e_{11,13}+e_{14,16}+e_{17,18}+e_{19,20}+e_{21,22};
\\
E_2= \mathbb{I}+e_{4,5}+e_{6,7}+e_{8,10}+e_{19,21}+e_{20,22}+e_{23,24};
\\
E_3= \mathbb{I}+e_{2,3}+e_{9,11}+e_{12,14}+e_{15,17}+e_{20,23}+e_{22,24};
\\
E_4= \mathbb{I}+e_{3,4}+e_{7,9}+e_{10,12}+e_{17,19}+e_{18,20}+e_{24,25};
\\
E_5= \mathbb{I}+e_{4,6}+e_{5,7}+e_{12,15}+e_{14,17}+e_{16,18}+e_{25,26};
\\
E_6= \mathbb{I}+e_{6,8}+e_{7,10}+e_{9,12}+e_{11,14}+e_{13,16}+e_{26,27}
\ee
together with the transposed matrices for $E_{\bar j}$, $j=1,\ldots,6$. The appropriate basis
for the Cartan generators $H_i=\exp{\left(h^i\log z \right)}, \ \ i=1,..,6$, is given by diagonal matices:
\be
h^1=\{\frac43, \underbrace{\frac13}_{11},  -\frac23, \underbrace{\frac13}_{2},\underbrace{ -\frac23,  \frac13}_{3},
   \underbrace{-\frac23}_{6}\},
\ \ \
h^2=\{\underbrace{1}_{4},\underbrace{0,1}_{2},\underbrace{0}_{12},\underbrace{-1}_{2},0,\underbrace{-1}_{4}\},
\\
h^3=\{\underbrace{\frac53}_{2}, \underbrace{\frac23}_{8}, -\frac13,  \frac23, \underbrace{-\frac13}_{2},\frac23 \underbrace{-\frac13}_{7}, -\underbrace{\frac43}_{5} \},
\ \ \
h^4= \{\underbrace{2}_{3},\underbrace{1}_{5},0,1,\underbrace{0}_{8},\underbrace{-1}_{6},\underbrace{-2}_{3}\},
\\
h^5= \{ \underbrace{\frac43}_{5}, \underbrace{\frac13}_{9},  -\frac23,  \frac13,  \underbrace{-\frac23}_{9},   \underbrace{-\frac53}_{2} \},
\ \ \
h^6= \{\underbrace{\frac23}_{7},\underbrace{-\frac13,\frac23}_{3},   \underbrace{-\frac13}_{13},  -\frac43 \}.
\ee
and multiplicities under the figure brackets denote the number of corresponding consequent combinations (if it is not unit), their sum is equal to the dimension of vector representation.

\begin{figure}[t]
\center{\includegraphics[width=4cm]{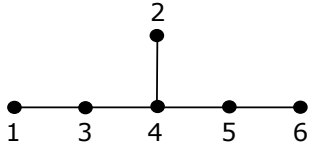}}
\caption{\sl Dynkin diagram for $E_6$.}
\label{fi:E6}
\end{figure}

The Lax equation \rf{RDE} now reads
\be
\label{LaxeqE6}
\det \left(g({\bf x},{\bf y}) + \mu\right) = \mu^{27} + \sum_{j=1}^{26}\mu^jR_j({\bf x},{\bf y}) + 1
\ee
and part of the coefficients of this characteristic polynomial can be identified immediately as
\be
\label{RrE6}
R_{26}=\Tr g = \Tr_{\bmu_{1}}g({\bf x},{\bf y}) = \mathcal{H}_1,
\\
R_1=\Tr g^{-1} =\Tr_{\bmu_6}g({\bf x},{\bf y}) = \mathcal{H}_6,
\\
R_{25}= \frac{1}{2}\left( \Tr g^2-(\Tr g)^2 \right) =\Tr_{\bmu_3}g({\bf x},{\bf y})= \mathcal{H}_3,
\\
R_3=\frac{1}{2}\left( \Tr g^{-2}-(\Tr g^{-1})^2 \right) =\Tr_{\bmu_5}g({\bf x},{\bf y})= \mathcal{H}_5,
\\
R_4=R_{24}=\frac13\Tr g^3+\frac16 (\Tr{g})^3-
\frac12\Tr g^2\Tr g =
\\
=\frac13\Tr g^{-3}+\frac16 (\Tr{g}^{-1})^3-\frac12\Tr g^{-2}\Tr g^{-1} =\Tr_{\bmu_{4}}g({\bf x},{\bf y})= \mathcal{H}_4
\ee
where the highest weights $\{\bmu_i\}$ correspond to the fundamental representations in the vertices of
the Dynkin diagram, as on fig.~\ref{fi:E6}. These formulas are again the consequences of the multiplication rules for the vector $V = V(\bmu_1)$ and its dual $V^\ast = V(\bmu_6)$ representations
\be
V(\bmu_3) = \wedge^2 V, \;\;
V(\bmu_5) = \wedge^2 V^\ast, \;\;
V(\bmu_4) = \wedge^3 V = \wedge^3 V^\ast
\ee
and for the $27\times 27$ matrices it is already technically easier to compute the traces of their lower powers instead of the characteristic polynomial, whose coefficients can be then calculated using the Newton formulas. However, expressions \rf{LaxeqE6} give directly only five integrals of motion out of six, which are necessary to ensure complete integrability: the vanishing of their Poisson bracket, determined by
blowup of the Dynkin diagram from fig.~\ref{fi:E6}.

To find the last ``minimal'' integral of motion let us use, that
\be
\wedge^4 V = V(\bmu_2+\bmu_5)
\ee
and the corresponding invariant function is
\be
\label{R23}
R_{23} =  \frac{1}{24}(\Tr g)^4 + \frac12 (\Tr g^2)^2 +
\frac13 (\Tr{g}^3)\Tr g -\frac14\Tr g^4 -\frac14\Tr g^2(\Tr g)^2 =
\\
= \Tr_{\wedge^4 V}g({\bf x},{\bf y}) = \Tr_{(\bmu_2+\bmu_5)}g({\bf x},{\bf y})
\ee
Using the following co-multiplication rules for $E_6$
\be
V(\bmu_2)\otimes V(\bmu_5) = V(\bmu_2+\bmu_5) + V(\bmu_3+\bmu_6) + V(\bmu_1+\bmu_2) +
V(2\bmu_6) + V(\bmu_5) + V(\bmu_1)
\\
V(\bmu_3)\otimes V(\bmu_6) = V(\bmu_3+\bmu_6) + V(\bmu_1+\bmu_2) + V(\bmu_5) + V(\bmu_1)
\\
V(\bmu_6)\otimes V(\bmu_6) = V(2\bmu_6) + V(\bmu_5) + V(\bmu_1)
\ee
one gets, that
\be
V(\bmu_2)\otimes V(\bmu_5) + V(\bmu_5) + V(\bmu_1) = V(\bmu_2+\bmu_5) + V(\bmu_3)\otimes V(\bmu_6) + V(\bmu_6)\otimes V(\bmu_6)
\ee
In terms of Ad-invariant functions it leads to the following relation
\be
\mathcal{H}_2\mathcal{H}_5 + \mathcal{H}_5 + \mathcal{H}_1 = R_{23} + \mathcal{H}_3\mathcal{H}_6+\mathcal{H}_6^2
\ee
which can be used to find
\be
\mathcal{H}_2 = \Tr_{\bmu_{2}}g({\bf x},{\bf y}) = {R_{23} + \mathcal{H}_3\mathcal{H}_6+\mathcal{H}_6^2 - \mathcal{H}_1\over \mathcal{H}_5}-1
\ee
in terms of already known \rf{RrE6} and \rf{R23}.

\section{Toda series for non simply-laced groups}

For the simple Lie groups of non ADE types one can construct the corresponding Toda lattices by
folding of the integrable systems, corresponding to the simply-laced groups. The folding procedure has
been proposed in \cite{DrinFG} for construction of the corresponding $\mathcal{X}$-cluster varieties, and it results in introducing of cluster coordinates on the symplectic leaves in non simply-laced groups, whose Poisson quivers arise from the gluing
of certain vertices for the Poisson graphs of symplectic leaves for the simply-laced groups.

For the non simply laced groups the cluster structure is generally described by enlarged set of data \cite{DrinFG},
which includes multipliers - the set of positive integers, all of them were equal to unities for
the symplectic leaves in the ADE cases, while generally these multipliers $d_i = \half(\alpha_i,\alpha_i)$
are determined by the lengths of the simple roots. Therefore, one has to distinguish here between the
Cartan matrix $C$ (or related with it cluster function $\epsilon$) and its symmetrization $B$ (related
with the skew-symmetric Poisson tensor $\hat{\epsilon}$), which gives directly the Poisson structure.
In this section we construct the Poisson graphs for the Coxeter-Toda symplectic leaves, which encode
the Poisson structures, given by $\hat{\epsilon}$, whose matrix elements are not necessarily integers,
in such cases we are going to use the dashed lines and put the corresponding values on them explicitly.

\begin{figure}[t]
\center{\includegraphics[width=17cm]{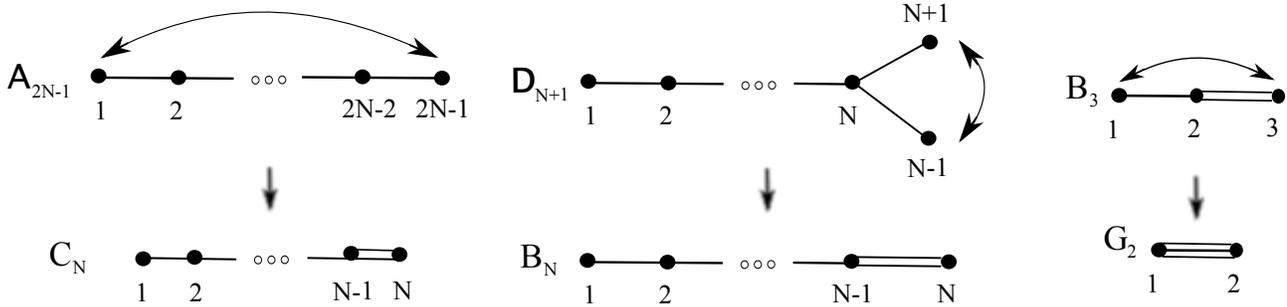}}
\caption{\sl Folding of Dynkin diagrams.}
\label{fi:Dynfold}
\end{figure}

On the level of Dynkin diagrams this folding be depicted as on fig.~\ref{fi:Dynfold}, which corresponds
to the following embedding:
\begin{itemize}
  \item $C_{N} \subset A_{2N-1}$, $N\geq 2$. The Dynkin diagram of $A$-series with odd number of vertices is folded twice, so that the simple root, corresponding to the middle vertex, remains untouched, while the other ones, not connected by a link on $A$-diagram, are pairwise summed.
  \item $B_{N} \subset D_{N+1}$, $N\geq 2$. Here all roots remain untouched by folding, except for
  two summed up, corresponding to the one-step ends of two links, growing out of the 3-valent vertex.
  Notice, that for $N=2$ this case coincides with the previous one.
  \item Exceptional cases: $F_4\subset D_5$ and $G_2 \subset B_3$. We shall consider below only the last one in sect.~\ref{ss:G_2}, remaining obtaining the $F_4$ system as an excercise for a reader.
\end{itemize}

The Poisson bracket \rf{pbN} turns after folding into the bracket
\be
\label{pbB}
\{ y_i,x_j\}_{\mathfrak{g}} = B_{ij}^{\mathfrak{g}}y_ix_j,\ \ \ \ i,j=1,\ldots,\rank\ G
\ee
where $B_{ij}^{\mathfrak{g}} = {1\over d_id_j}(\alpha_i,\alpha_j)$ is the symmetrized
Cartan matrix for the non simply-laced case \cite{Kac}.

The rest to check is that after folding the integrals of motion of the original simply-laced
Toda system turn into the integrals of motion of new integrable system, i.e. they are still
in the involution w.r.t. the new Poisson structure, and one still have enough independent
integrals of motion to ensure complete integrability. Counting of the number of integrals of motion
is still very simple: each gluing of
two vertices on a Dynkin diagram reduces rank of the group by one, i.e. reduces by one the
total amount of the Ad-invariant functions. 	
Simultaneously this corresponds to gluing two vertices on the Poisson quiver, i.e. reduces the
dimension of the phase space by two.

The fact, that all integrals of motion are still in the involution is almost obvious from
general reasoning: after folding one gets a symplectic leaf in the group, which is embedded
as a subgroup into original simply-laced group. The Poisson structure on symplectic leaf is
according to \cite{DrinFG} a restriction of the $r$-matrix bracket \rf{rbra}, so that any two
Ad-invariant functions are in the involution. By construction this is the property of our integrals
of motion, which is preserved after reduction, and taking into account proper counting one concludes,
that we still have an integrable system. We are now going to illustrate this on particular examples.

\subsection{$B_2=C_2$ example
\label{ss:B2}}

This case can be obtained by reduction of the $SL(4)$ relativistic Toda in the following way: one embeds $C_2\subset A_3=D_3$ (top left on fig.~\ref{fi:Dynfold} for $N=2$) by
\be
\label{emB2A3}
e_1 = e_1'+e_3',\ \ \ \ \  e_2=e_2'
\\
f_1 = f_1'+f_3',\ \ \ \ \ f_2=f_2'
\\
h_1 = h_1'+h_3',\ \ \ \ \ h_2=h_2'
\ee
where $\{e,f,h\}$ and $\{e',f',h'\}$ are the Chevalley bases of $C_2$ and $A_3$ correspondingly (see notations in Appendix~\ref{ap:promunu}). An essential point here is that $C'_{13}=C'_{31}=0$ for
$SL(4)$, so that the generators \rf{emB2A3} satisfy the canonical commutational relations \rf{rela} with the Cartan matrix
\be
\label{CB2C2}
C = \|C_{ij}\| = \left(\begin{array}{cc}
                         2 & -2 \\
                         -1 & 2
                       \end{array}\right)
\ee
This Cartan matrix is already not symmetric, since
$[h_1,e_2]=[h_1'+h_3',e_2'] = -2e_2'=-2e_2$, while $[h_2,e_1]=[h_2,e_1'+e_3'] = -e_1'-e_3'=-e_1$,
but it can be symmetrized (see e.g. \cite{Kac}) by
\be
\label{BABAnum}
C = \left(\begin{array}{cc}
                         2 & -2 \\
                         -1 & 2
                       \end{array}\right) =
                       \left(\begin{array}{cc}
                         2 & 0 \\
                         0 & 1
                       \end{array}\right)\left(\begin{array}{cc}
                         1 & -1 \\
                         -1 & 2
                       \end{array}\right)
\ee
or
\be
\label{BABA}
\|C_{ij}\| = \|d_iB_{ij} \|,\ \ \ \ \ d_i = {(\alpha_i,\alpha_i)\over 2}
\\
\|B_{ij}\| = {1\over d_id_j}\|(\alpha_i,\alpha_j)\| = \left(\begin{array}{cc}
                         1 & -1 \\
                         -1 & 2
                       \end{array}\right)
\ee
which is just matrix of the normalized scalar products of two simple roots for $B_2=C_2$.

\begin{figure}[hc]
\center{\includegraphics[width=75pt]{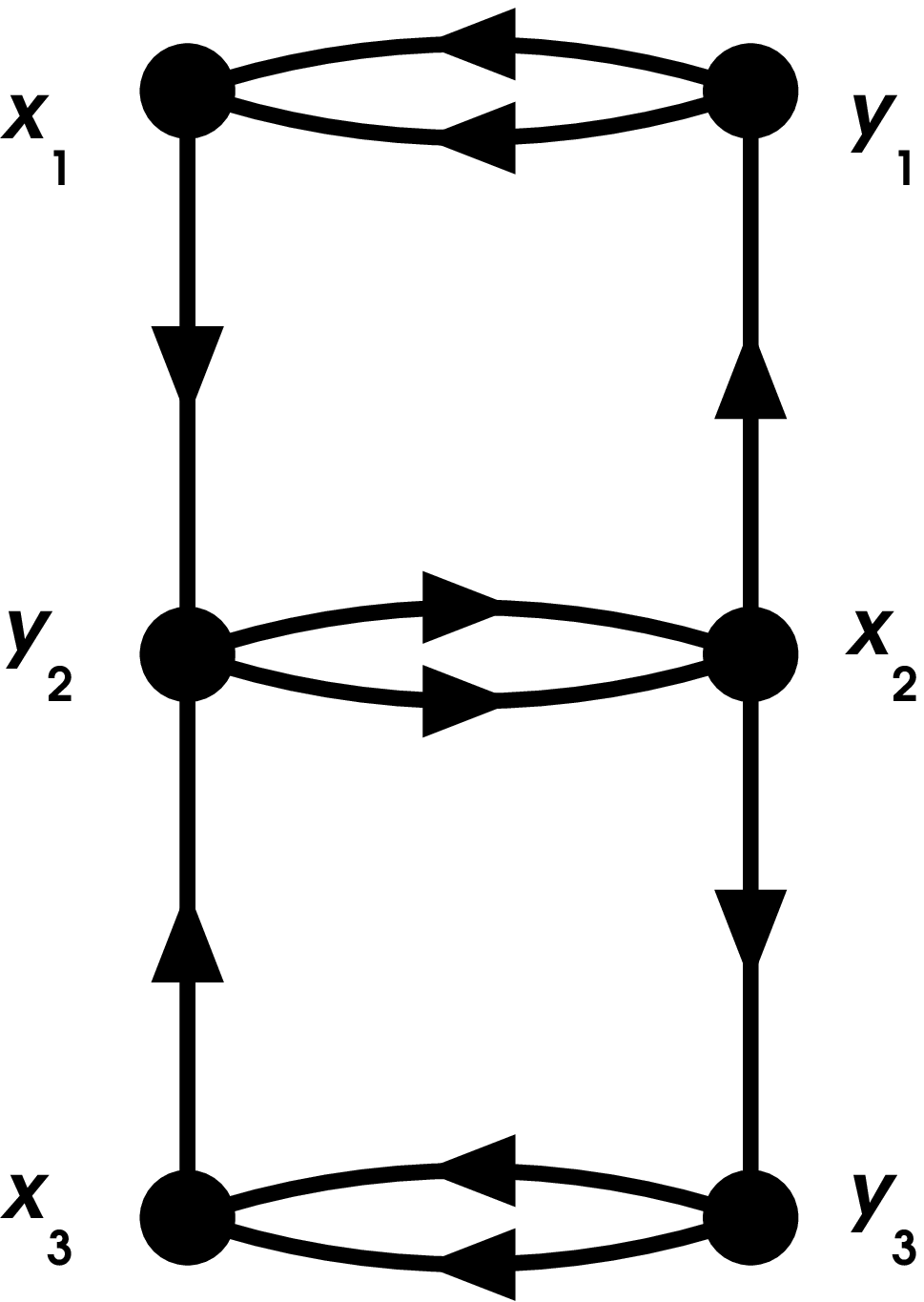}\hspace{1cm}
\includegraphics[width=130pt]{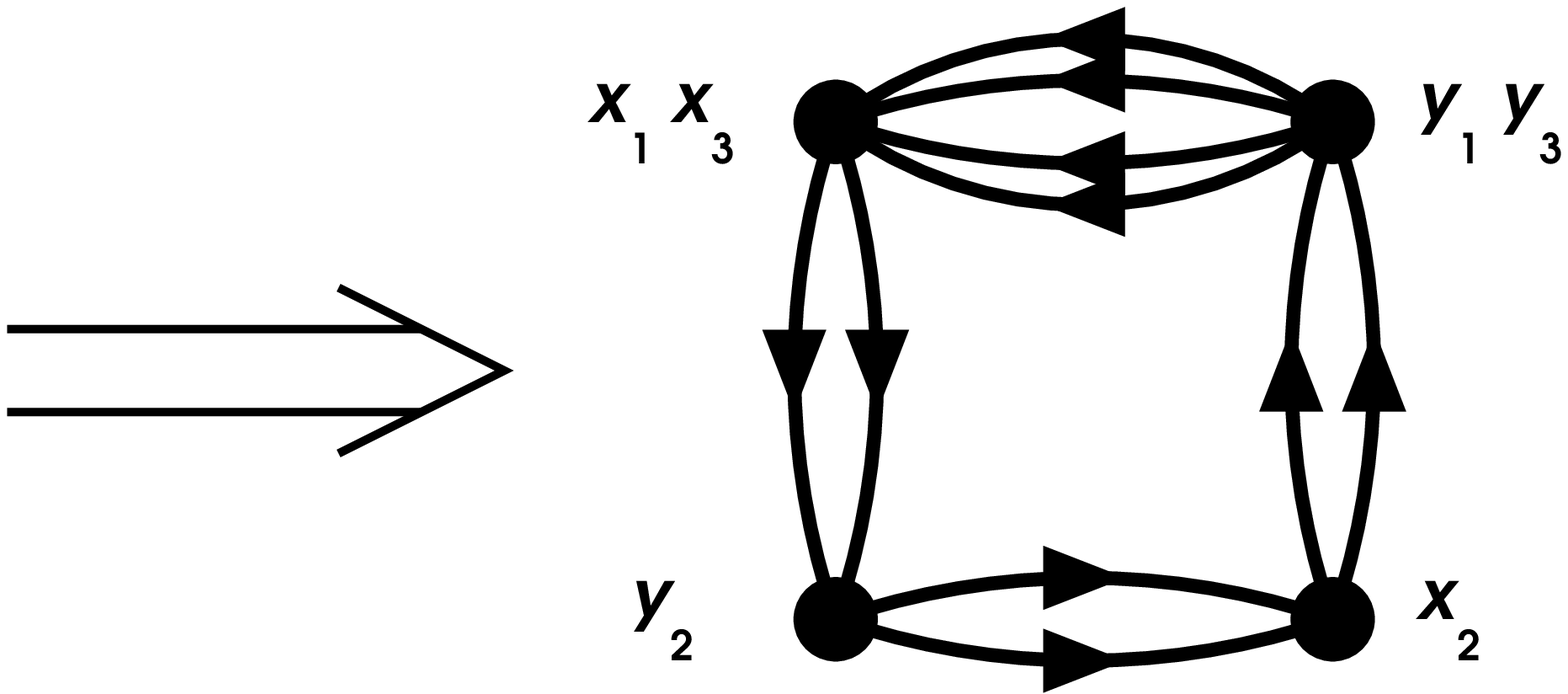}\hspace{1cm}
\includegraphics[width=120pt]{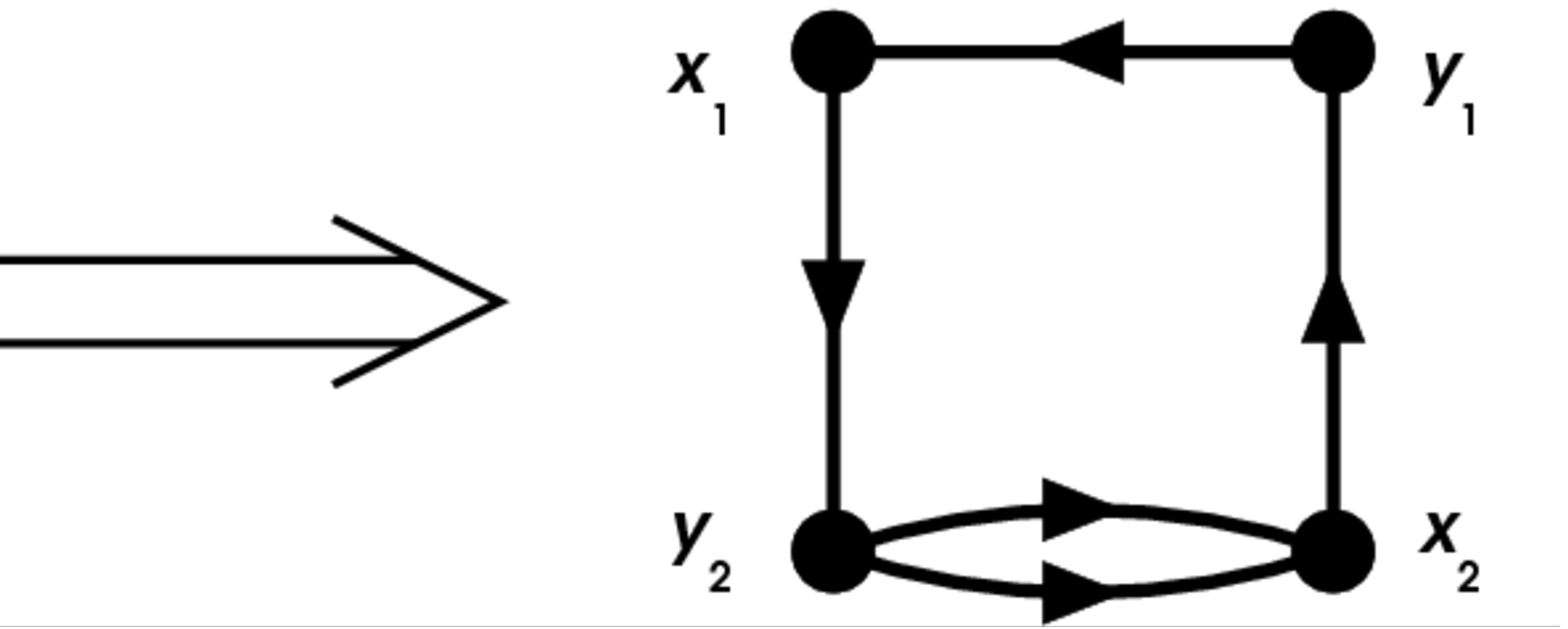}}
\caption{\sl Construction of a graph for the Toda symplectic leave in $C_2$ from $SL(4)$.
Gluing the left picture one gets the graph in the middle with the products $x_1x_3$ and $y_1y_3$ in
the upper vertices, which become $x_1^2$ and $y_1^2$ after identification $x_1=x_3$ and $y_1=y_3$ corresponding to \rf{emB2A3}. It means that new variables are coordinates on the cluster variety,
corresponding to the right graph, with the exchange matrix generated by $B$ from \rf{BABA}.}
\label{fi:A3B2}
\end{figure}

The Poisson structure on symplectic leaves of $B_2$ of dimension $2 \cdot \mbox{rank}=4$ can be
obtained from the induced folding \cite{DrinFG} of the Coxeter-Toda cluster variety in $A_3=SL(4)$ of dimension 6.
The most transparent way to get it comes from gluing the corresponding vertices on the Poisson graph, the corresponding procedure is depicted on fig.~\ref{fi:A3B2}.
It results in four dimensional symplectic leaf in $B_2$ or $C_2$, where the cluster variables
have the following the Poisson brackets
\be
\{ y_i,x_j\}_{B_2} = B_{ij} y_ix_j,\ \ \ i,j=1,2
\\
\{ x_i,x_j\}_{B_2} =\{ y_i,y_j\}_{B_2} =0
\ee
i.e. encoded by matrix from \rf{BABA}.

The Lax operator
\be
\label{gaussB2}
g({\bf x},{\bf y})  \simeq
\underbrace{H_1(y_1)E_1H_1(x_1)E_{\bar 1}}_{g_1'}\cdot
\underbrace{H_2(y_2)E_2H_2(x_2)E_{\bar 2}}_{g_2'}\cdot\underbrace{H_3(y_1)E_3H_3(x_1)E_{\bar 3}}_{g_3'}\simeq
\\
\simeq\underbrace{H_1(y_1)H_3(y_1)E_1E_3 \cdot H_1(x_1)H_3(x_1)E_{\bar 1}E_{\bar 3}}_{g_1} \cdot \underbrace{H_2(y_2)E_2H_2(x_2)E_{\bar 2}}_{g_2}
\ee
is just obtained by identification $x_1=x_3$ and $y_1=y_3$ in the formula \rf{gADE} for $SL(4)$, here $\simeq$ as usual means, that we consider equalities modulo cyclic permutation, which does not influence
Ad-invariant functions. For the characteristic polynomial of \rf{gaussB2} one gets
\be
\label{RB2}
\det \left(\mu + g({\bf x},{\bf y})\right) =
\mu^4 +  \mathcal{H}_1({\bf x},{\bf y})\mu^3 + \mathcal{H}_2({\bf x},{\bf y})\mu^2 + \mathcal{H}_1({\bf x},{\bf y})\mu +1
\ee
with only two independent integrals of motion
\be
\label{hamB2}
\mathcal{H}_1({\bf x},{\bf y}) = \frac{1}{x_1y_1\sqrt{x_2}\sqrt{y_2}}(y_1^2y_2x_1^2x_2+y_1^2y_2x_1x_2+y_1y_2x_1x_2+y_1y_2x_1+y_1x_1+y_1+1)
\\
\mathcal{H}_2({\bf x},{\bf y}) = \frac{1}{x_1y_1x_2y_2}(y_1^2y_2^2x_1^2x_2^2+y_1^2y_2^2x_1^2x_2+y_1^2y_2x_1^2x_2+2y_1^2y_2x_1x_2+y_1^2y_2x_2+
\\
2y_1y_2x_1x_2+2y_1y_2x_2+y_2x_2+y_2+1)
\ee
It is easy to check, that $\{ \mathcal{H}_1({\bf x},{\bf y}),\mathcal{H}_2({\bf x},{\bf y})\}_B=0$, as
should follow from the proposed folding construction.

The Darboux coordinates in this case are given by modification of the standard simply-laced formulas
\rf{darbun}
\be
\label{darbuB2}
x_1 = e^{-q_1},\ \ \ x_2=e^{q_1-q_2}
\\
y_1=e^{p_1+q_1}f_1(q),\ \ \ y_2=e^{p_2-p_1+q_2-q_1}f_2(q)
\ee
with the ``long momentum'' \rf{Pp}, generated by
\be
\label{SB2}
S(q) = \sum_{k=1,2}{1\over d_k}{\rm Li}_2\left(-\exp(\alpha_k\cdot q)\right) =
{\rm Li}_2\left(-e^{q_1}\right)+\half {\rm Li}_2\left(-e^{q_2-q_1}\right)
\ee
i.e. $f_i={\d S\over\d q_i}$, $i=1,2$ in \rf{darbuB2}. In these variables the first Hamiltonian in
\rf{hamB2} becomes
\be
\mathcal{H}_1 = \left(e^{p_1+p_2\over 2} + e^{-{p_1+p_2\over 2}}\right)\sqrt{1+e^{q_1}}
+  \left(e^{p_1-p_2\over 2} + e^{-{p_1-p_2\over 2}}\right)\sqrt{1+e^{q_1}}\sqrt{1+e^{q_2-q_1}}
\ee
In the limit to Lie algebra this turns into
\be
\mathcal{H}_1 \approx \left({p_1+p_2\over 2}\right)^2 + \left({p_1-p_2\over 2}\right)^2
+ 2e^{q_1} + e^{q_2-q_1} =
\half(p_1^2+p_2^2) + 2e^{q_1} + e^{q_2-q_1}
\ee
the Hamiltonian of $B_2=C_2$ open Toda lattice. This procedure can be performed for the other groups as well, but we remain it to a reader as an exercise.

\subsection{$B_3$ versus $C_3$ systems
\label{ss:B3C3}}

\begin{figure}[hc]
\center{\includegraphics[width=14cm]{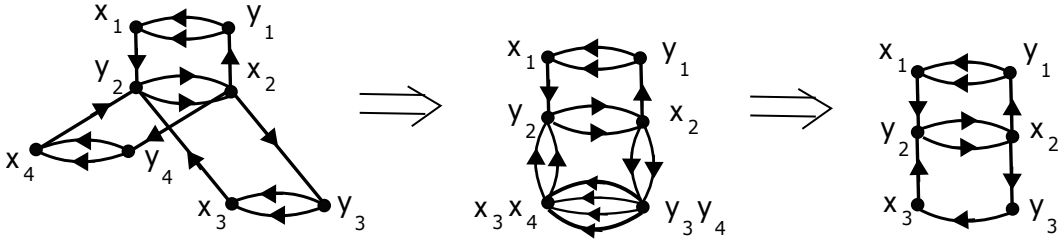}}
\caption{\sl Construction of a graph for $B_3=SO(7)$ by reduction from $D_4$.}
\label{fi:todaso7}
\end{figure}
In this example one considers the reduction from $D_4$ (see fig.~\ref{fi:todaso7}), corresponding to gluing two from three vertices with further identification of the cluster variables, for example
\be
x_3=x_4, \ \ y_3=y_4
\ee
It gives rise to explicit formulas for the integrals of motion: two of them are obviously constructed
from the function \rf{rfun}, these are
\be
\mathcal{H}_1=\frac{1}{y_1x_1y_2x_2y_3x_3}(y_3^2y_2^2y_1^2x_3^2x_2^2x_1^2+y_3^2y_2^2y_1^2x_3^2x_2^2x_1+y_3^2y_2^2y_1x_3^2x_2^2x_1+y_3^2y_2^2y_1x_3^2x_2x_1+
\\
y_3^2y_2y_1x_3^2x_2x_1+2y_3^2y_2y_1x_3x_2x_1+y_3^2y_2y_1x_2x_1+2y_3y_2y_1x_3x_2x_1+2y_3y_2y_1x_2x_1+y_2y_1x_2x_1+
\\
y_2y_1x_1+y_1x_1+y_1+1) =
\mathfrak{h}(x_1,y_1|x_3,y_3;x_3,y_3|x_2,y_2)
\ee
and
\be
\mathcal{H}_3=\frac{1}{y_3^{3/2}x_3^{3/2}y_2x_2\sqrt{y_1x_1}}(y_3^3y_2^2y_1x_3^3x_2^2x_1+y_3^3y_2^2y_1x_3^2x_2^2x_1+y_3^2y_2^2y_1x_3^2x_2^2x_1+y_3^2y_2^2y_1x_3^2x_2x_1+
\\
y_3^2y_2y_1x_3^2x_2x_1+y_3^2y_2y_1x_3^2x_2+y_3^2y_2y_1x_3x_2x_1+y_3^2y_2y_1x_3x_2+y_3^2y_2x_3^2x_2+y_3y_2y_1x_3x_2x_1+
\\
y_3^2y_2x_3x_2+y_3y_2y_1x_3x_2+y_3y_2x_3x_2+
y_3y_2x_3+y_3x_3+y_3+1) =
\\
= \mathfrak{h}(x_3,y_3|x_1,y_1;x_3,y_3|x_2,y_2)
\ee
The third integral, which corresponds to the middle vertex on the Dynkin diagram is obtained by the
same procedure, and its explicit form is presented in Appendix~\ref{ap:explicit}, formula \rf{R2B3}.
These integrals of motion are in involution with respect to the Poisson structure,
read from the right graph at fig.~\ref{fi:todaso7}.

\begin{figure}[hc]
\center{\includegraphics[width=12cm]{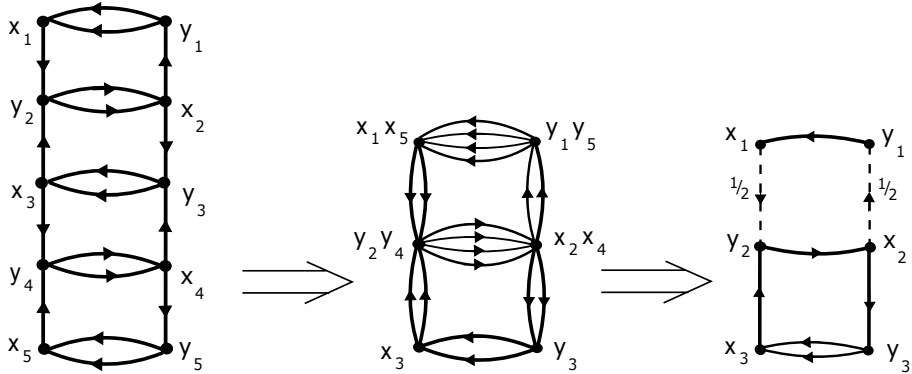}}
\caption{\sl Construction of a graph for $C_3=SP(7)$.}
\label{fi:todasp7}
\end{figure}
In the $C_3$ case the reduction comes instead from the $A_5=SL(6)$ theory, where it reads
\be
x_1=x_5, \ \ x_2=x_4, \ \ y_1=y_5, \ \ y_2=y_4
\ee
which corresponds to embedding
\be
e_1=e'_1+e'_5, \ \ \ \ f_1=f'_1+f'_5
\\
e_2=e'_2+e'_4, \ \ \ \ f_2=f'_2+f'_4
\\
e_3=e'_3
\ee
Integrals of motion are directly found from the coefficients of characteristic polynomial,
so that the simplest one is:
\be
\label{C3first}
\mathcal{H}_1=\mathcal{H}_5=\frac{1}{y_1y_2 \sqrt{y_3} x_1x_2 \sqrt{x_3}} ( y_1^2y_2^2y_3x_1^2x_2^2x_3+y_1^2y_2^2y_3x_1x_2^2x_3+y_1y_2^2y_3x_1x_2^2x_3+y_1y_2^2y_3x_1x_2x_3+
\\
y_1y_2y_3x_1x_2x_3+y_1y_2y_3x_1x_2+y_1y_2x_1x_2+y_1y_2x_1+y_1x_1+y_1+1 )
\ee
The explicit form for second and third integrals of motion is presented in Appendix~\ref{ap:explicit}, formula \rf{R23C3}

\subsection{Toda system for $G_2$ group
\label{ss:G_2}}

\begin{figure}[hc]
\center{\includegraphics[width=14cm]{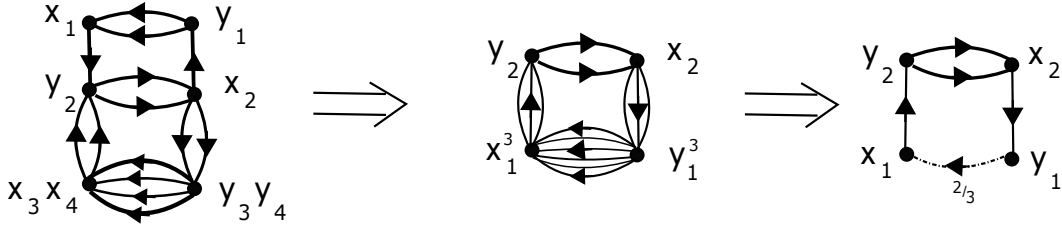}}
\caption{\sl Construction of the Poisson quiver for $G_2$.}
\label{fi:todaG2}
\end{figure}
Here one can use further reduction compare to what has been done in the $B_3$ case, or just put in the formulas for $D_4$:
\be
x_1=x_3=x_4,
\\
y_1=y_3=y_4
\ee
in the $D_4$ case. The resulting Poisson quiver (see fig.~\ref{fi:todaG2}) corresponds
to the right matrix
\be
\label{BABAG2}
C=\left(\begin{array}{cc}
                         2 & -3 \\
                         -1 & 2
                       \end{array}\right) =
                       \left(\begin{array}{cc}
                         3 & 0 \\
                         0 & 1
                       \end{array}\right)\left(\begin{array}{cc}
                         2/3 & -1 \\
                         -1 & 2
                       \end{array}\right) = D\cdot B
\ee
in the symmetrization decomposition of the Cartan matric for $G_2$.

This finally results in the Toda like integrable system with two integrals of motion
\be
\mathcal{H}_1=\frac{1}{y_1^2y_2x_1^2x_2}(y_1^4y_2^2x_1^4x_2^2+y_1^4y_2^2x_1^3x_2^2+y_1^3y_2^2x_1^3x_2^2+y_1^3y_2^2x_1^3x_2+y_1^3y_2x_1^3x_2+2y_1^3y_2x_1^2x_2+
\\
+y_1^3y_2x_1x_2+2y_1^2y_2x_1^2x_2+2y_1^2y_2x_1x_2+y_1y_2x_1x_2+y_1y_2x_1+y_1x_1+y_1+1)
\\
\mathcal{H}_2=\frac{1}{y_1^3y_2^2x_1^3x_2^2}(y_1^6 y_2^4 x_1^6 x_2^4+y_1^6 y_2^4 x_1^6 x_2^3+y_1^6 y_2^3 x_1^6 x_2^3+3 y_1^6 y_2^3 x_1^5 x_2^3+3 y_1^6 y_2^3 x_1^4 x_2^3+3 y_1^5 y_2^3 x_1^5 x_2^3+
\\
+y_1^6 y_2^3 x_1^3 x_2^3+6 y_1^5 y_2^3 x_1^4 x_2^3+3 y_1^5 y_2^3 x_1^3 x_2^3+3 y_1^4 y_2^3 x_1^4 x_2^3+3 y_1^4 y_2^3 x_1^4 x_2^2+3 y_1^4 y_2^3 x_1^3 x_2^3+3 y_1^4 y_2^3 x_1^3 x_2^2+
\\
+3 y_1^4 y_2^2 x_1^4 x_2^2+y_1^3 y_2^3 x_1^3 x_2^3+6 y_1^4 y_2^2 x_1^3 x_2^2+2 y_1^3 y_2^3 x_1^3 x_2^2+3 y_1^4 y_2^2 x_1^2 x_2^2+y_1^3 y_2^3 x_1^3 x_2+4y_1^3y_2^2x_1^3x_2^2+
\\
+2y_1^3y_2^2x_1^3x_2+6y_1^3y_2^2x_1^2x_2^2+3y_1^3y_2^2x_1^2x_2+y_1^3y_2x_1^3x_2+3y_1^2y_2^2x_1^2x_2^2+3y_1^3y_2x_1^2x_2+3y_1^2y_2^2x_1^2x_2+
\\
+3y_1^3y_2x_1x_2+3y_1^2y_2x_1^2x_2+y_1^3y_2x_2+6y_1^2y_2x_1x_2+
3y_1^2y_2x_2+3y_1y_2x_1x_2+
\\
3y_1y_2x_2+y_2x_2+y_2+1)
\ee
which are in involution w.r.t. the Poisson structure given by right graph from fig.~\ref{fi:todaG2}.

\section{Affine Lie groups}

The cluster construction of symplectic leaves in simple Lie groups has nice generalization to the affine $\hat{A}$-series \cite{AMJGP,FM}, which gives rise, in addition with affine Toda chains, to a large class of integrable models, coinciding with recently described by Goncharov and Kenyon \cite{GK} in terms of
the dimer models on bipartite graphs on two-dimensional torus. The new important features of the
affine generalization are:
\begin{itemize}
  \item the Poisson submanifolds are naturally constructed only for the \emph{co-extended} affine groups $\widehat{G}^\sharp=\widehat{G}\rtimes \mathbb{C}^\times$ (in contrast to more common
      their central extension $\widehat{G}\ltimes \mathbb{C}^\times$), see \cite{Kac},
      with final projection to the elements with trivial co-extension. In practice one has to extend the Cartan subalgebra by the gradation or derivative (in spectral parameter) operator $h^0=\lambda{\d\over\d\lambda}$, which has nontrivial commutational relations with other generators of Lie algebra, or to add its exponent $T_z=z^{\lambda{\d\over\d\lambda}}$ with $z\in \mathbb{C}^\times$ to the Cartan subgroup of group $\widehat{G}$. For the groups $\widehat{PGL}^\sharp(N)$ the Poisson submanifolds are then labeled by the cyclically irreducible elements of the co-extended double Weyl group, the simple roots for this group can be identified with the set $\Pi=\mathbb{Z}/N\mathbb{Z}$, with the Dynkin diagram given by a closed necklace with $N$ vertices, and its co-extension adds a discrete rotational automorphism $\Lambda$ of this
      diagram, satisfying $\Lambda^N=1$, see \cite{FM};
  \item the Lax operators, constructed as in \rf{gADE} taking products, parameterized by words in  co-extended double Weyl group, depend on spectral parameter, introduced in a natural way by taking extra affine root generators in the evaluation representation. The characteristic polynomial equation gives rise to a spectral curve, embedded in $C^\times\times C^\times$, where the
      coefficients are naively defined ambiguously, but certain invariant combinations of them still produce the full set of integrals of motion.
\end{itemize}

The detailed investigation of these issues for other than $\hat{A}$-type affine Lie groups and
their connection with dimer models go beyond the scope of this paper. However, we are going now
to present the results for the spectral curves of affine Toda chains of $\widehat{D}_N$ family, and to discuss in detail a particular example of ${\widehat{D_4}}$ relativistic Toda model, which demonstrates, that generalization in fact works in almost straightforward way.

\subsection{${\hat{D_4}}$ example
\label{ss:{hatD_4}}}

\begin{figure}[hc]
\center{\includegraphics[width=9cm]{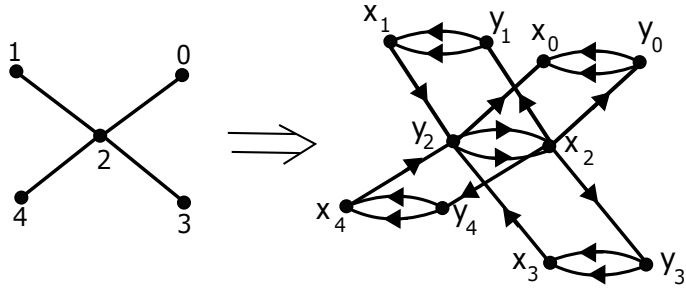}}
\caption{\sl Blowup of Dynkin diagram and corresponding Poisson quiver in $\hat{D_4}$. It can be
constructed from gluing Poisson subgroups, corresponding to the simple roots, just in the same way, as on fig.~\ref{fi:todaD4}.}
\label{fi:hatD4}
\end{figure}
For the Poisson submanifolds in affine Lie groups the Poisson structure is degenerate. In the case of Toda submanifold in the affine ${\widehat{D_4}}$ it can be viewed again as a blowup of
the corresponding Dynkin diagram (see fig.\ref{fi:hatD4}), and it is easy to see, that
\be
\label{casimir}
\mathcal{C}=x_0x_1x_2^2x_3x_4=(y_0y_1y_2^2y_3y_4)^{-1}
\ee
is a Casimir function. The last equality here can be interpreted as condition of trivial total co-extension, already mentioned above and arising here with a slight modification from \cite{AMJGP,FM}.

For the co-extended ${\widehat{D_4}}^\sharp$ one can choose the following basis of the generators:
\be
\label{HDh}
\hat{H_i}(z)=H_i(z)T_{z^{\gamma_i}}, \;\; \text{where} \;\; \gamma_{1,3,4}=1 \;\; \text{and} \;\; \gamma_2=2
\\
\hat{H}_0(z)=T_z=z^{\lambda\frac{\partial}{\partial \lambda}}=\exp \left( \log z\cdot\lambda\frac{\partial}{\partial \lambda} \right)
\ee
for the Cartan part, where $T_{z}$ is an operator of (multiplicative) translation, with nontrivial exchange relations $T_{z}f(\lambda)= f(z\lambda)T_z$ with each $\lambda$-depended extra affine element, and
\be
\label{EDh}
E_i = \exp(e_i),\ \ \ \ E_{\bar i} = \exp(e_{\bar i}),\ \ \ \ i,{\bar i}=1,\ldots,N=4
\\
E_0(\lambda)=\exp\left( \lambda e_{\bar 0} \right) , \; E_{\bar 0}(\lambda)=\exp\left( \lambda^{-1} e_0 \right)
\ee
for the exponentiated simple root generators. An arbitrary element of ${\hat{D_4}}^\sharp$ can be constructed by taking all possible products of generators \rf{HDh} and \rf{EDh}.

The additional affine simple root generator corresponds to the highest root of $D_4$, and can be found
explicitly, using \rf{ED4}:
\be
e_0=\left[ \left[ \left[ [e_1,e_2],e_3\right],e_4\right]e_2\right]=\left[ \left[ \left[ [E_1,E_2],E_3\right],E_4\right]E_2\right],\ \ \ \ e_{\bar 0}=e_0^{\rm tr}
\\
E_{0}(\lambda) = \mathbb{I}+\lambda( e_{71}- e_{82}),\ \ \ \ \
E_{\bar 0}(\lambda) = \mathbb{I}+\lambda^{-1}(e_{17}-e_{28})
\ee
which follows, for example, from decomposition of the highest root (see \cite{prog}).

Now similar to (\ref{gADE}) formula defines
\be
\hat{g}\simeq\prod_{i=0}^4\hat{H_i}(y_i)E_i\hat{H_i}(x_i)E_{\bar i} H_0(y_0)E_0(\lambda)H_0(x_0)E_{\bar 0}(\lambda) \simeq
\\
\simeq\prod_{i=0}^4H_i(y_i)E_i H_i(x_i)E_{\bar i} T_{y_1x_1y_2^2x_2^2y_3x_3y_4x_4} E_0\left( \lambda/x_{0} \right) E_{\bar 0}(\lambda)
\ee
an element of ${\hat{D_4}}^\sharp$, corresponding to the Toda Poisson submanifold. In order to diagonalize it one has to project
to a trivial co-extension $T_{y_1x_1y_2^2x_2^2y_3x_3y_4x_4}=\mathbb{I}$  or $y_1x_1y_2^2x_2^2y_3x_3y_4x_4=1$ (cf. now with right equality in \rf{casimir}), giving rise
to the spectral parameter dependent Lax operator
\be
\label{LaxD4aff}
g(\lambda)=\prod_{i=0}^4H_i(y_i)E_i H_i(x_i)E_{\bar i} E_0\left( \lambda/x_{0} \right) E_{\bar 0}(\lambda)
\ee
and reproducing
the last equality in \rf{casimir}.

\begin{figure}[hc]
\center{\includegraphics[width=8cm]{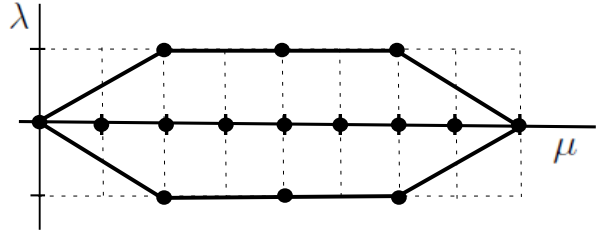}}
\caption{ \sl Newton polygon $\Delta$ for the spectral curve of $\hat{D_4}$. The set of its coefficients is symmetric under reflection w.r.t. the vertical line, passing through the center.}
\label{fi:NpolygonD4}
\end{figure}

Vanishing of the characteristic polynomial for \rf{LaxD4aff}
\be
F(\lambda, \mu)=\det \left( \hat{g} + \mu \cdot \mathbb{I} \right) = 0
\ee
gives the equation of the curve with
\be
\label{D4sc}
F(\lambda, \mu)=\sum_{i,j\in\Delta}Q_{ij}\lambda^{i}\mu^{j}=\mu^8+Q_{07}\mu^7 + \left( Q_{16}\lambda+Q_{06}+Q_{-16}\lambda^{-1} \right)\mu^6+Q_{05}\mu^5+
\\
+\left( Q_{14}\lambda+Q_{04}+Q_{-14}\lambda^{-1} \right)\mu^4+Q_{03}\mu^3+\left( Q_{12}\lambda+Q_{02}+Q_{-12}\lambda^{-1} \right)\mu^2+Q_{01}\mu +1 
\ee
whose Newton polygon is depicted at fig.~\ref{fi:NpolygonD4}. Its coefficients are defined
ambiguously (see detailed discussion of this issue in \cite{FM,AMJGP}) unless one redefines
the spectral parameters and total normalization of $F(\lambda, \mu)$ to make three boundary coefficients
to be unities.

For the equation \rf{D4sc} two such coefficients (at the leftmost and rightmost vertices) are
already unit, so it is enough to rescale $\lambda$ in order to make also unit one of the
coefficients $Q_{ij}$ for $i \neq 0$. It is easy to check, that then the rest $Q_{ij}$ with $i \neq 0$
are all expressed through the Casimir function $\mathcal{C}$, while the coefficients $Q_{0j}$ are integrals of motion, i.e.
\be
\{ Q_{1j}, x_k\} = \{ Q_{-1j}, x_k\} = \{ Q_{1j}, y_k\} = \{ Q_{-1j}, y_k\} = 0,\ \ \ \forall\ j,k
\\
\{ Q_{0i}, Q_{0j}\} = 0,\ \ \ \forall\ i,j
\ee
Due to the symmetry $Q_{0,j}=Q_{0,8-j}$, there are only four nontrivial independent integrals
of motion in this system.

Using constraints \rf{casimir} one can express
\be
x_0= \frac{\mathcal{C}}{x_1x_3x_4x_2^2},\ \ \ y_0= \frac{1}{\mathcal{C}y_1y_3y_4y_2^2}
\ee
and the rest independent variables parameterize the symplectic manifold, isomorphic to symplectic leaf in simple $D_4$, whose Poisson structure is given by the quiver at right picture from fig.\ref{fi:todaD4}. In terms of these independent variables one gets, for example, for
\be
Q_{01}=\frac{1}{y_1x_1y_2x_2\sqrt{y_3x_3}\sqrt{y_4x_4}}( 1+y_1+y_1x_1+y_1x_1y_2+ y_1x_1y_2x_2+y_1x_1y_2x_2y_3+
\\
+y_1x_1y_2x_2y_4
+ y_1x_1y_2x_2y_3x_3+y_1x_1y_2x_2y_4x_4+y_1x_1y_2x_2y_3y_4+ y_1x_1y_2x_2y_3x_3y_4+
\\
+y_1x_1y_2x_2y_4x_4y_3+
 y_1x_1y_2x_2y_3x_3y_4x_4+y_1x_1y_2^2x_2y_3x_3y_4x_4
 +y_1x_1y_2^2x_2^2y_3x_3y_4x_4+
 \\
 +y_1^2x_1y_2^2x_2^2y_3x_3y_4x_4+y_1^2x_1^2y_2^2x_2^2y_3x_3y_4x_4
+\frac{1}{\mathcal{C}}(x_1^2x_2^2x_3x_4y_1+x_1x_2^2x_3x_4y_1+x_1x_2^2x_3x_4))
\ee
In the limit $\mathcal{C} \rightarrow \infty$ one finds, that it exactly coincides with the first
integral of motion (\ref{rfun}) in relativistic Toda of simple group $D_4$. Similarly
in this limit one gets for all $Q_{0,i}=R_i$, where $R_i$ are coefficients of the characteristic
polynomial \rf{LaxeqD4}.

\begin{figure}[hc]
\center{\includegraphics[width=14cm]{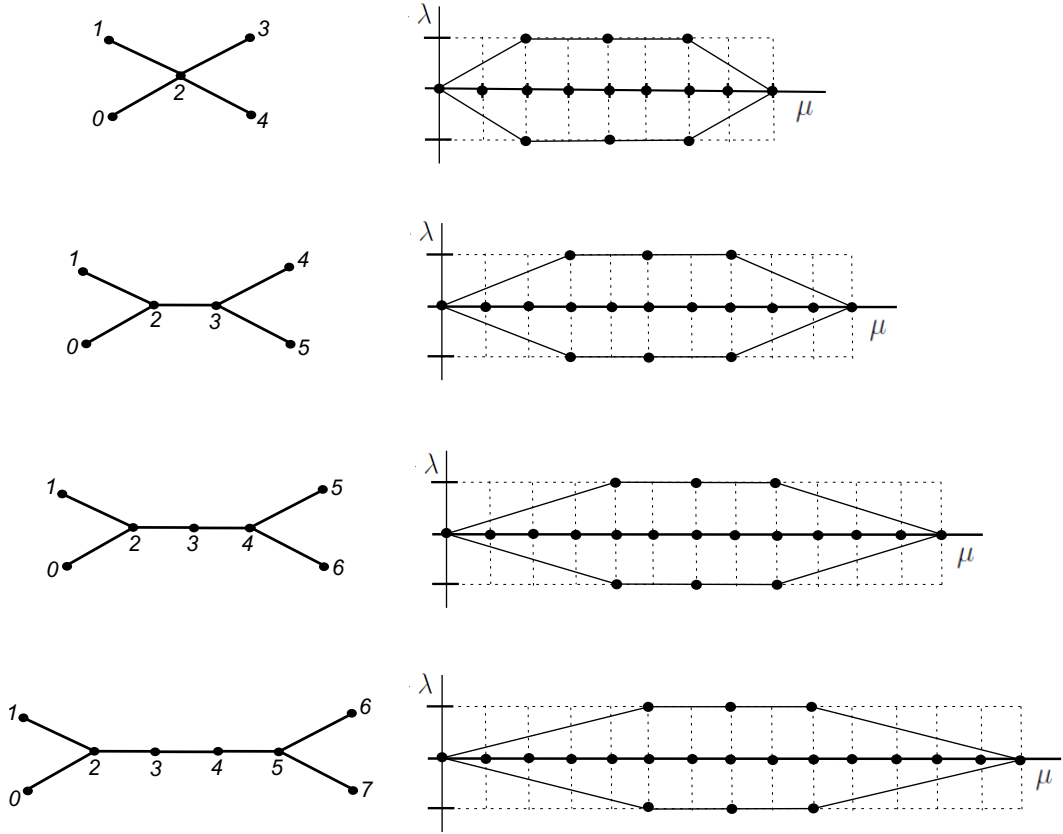}}
\caption{\sl Newton polygons $\Delta$ for the spectral curves of $\hat{D_n}$-series. The set of its coefficients is symmetric under reflection w.r.t. the vertical line, passing through the center.}
\label{fi:NpolygonDhat}
\end{figure}

\section{Conclusion}

In this paper we have extended the procedure of constructing of the integrable systems on the symplectic
leaves in simple Lie groups of appropriate dimension beyond the well-known relativistic Toda chains,
corresponding to A-series. We have shown, that for the simply-laced case the Poisson structures are
still described by Poisson quivers, directly obtained by blowing up the corresponding Dynkin
diagram, and proposed a procedure to get a set of minimal integrals of motion.

For the non simply-laced case the corresponding integrable models are obtained by folding of
the ADE-systems. We have considered this procedure in detail on several important examples
and found explicit formulas for their integrals of motion in cluster coordinates.

These integrals of motion generate continuous Hamiltonian flows in integrable models, which are linearized
on (degenerate) Jacobians of their spectral curves.
A separate interesting problem is to study the discrete flows in cluster integrable systems, which are
generated by the cluster mutations \cite{GSW,pent,SolKhe,FM}, and we postpone it for separate independent study.

The most interesting is the affine case, which already in the case of A-series gives rise
to a large class of integrable systems, which can be obtained alternatively by studying dimer models
on bipartite graphs on a two-dimensional torus \cite{GK}. The detailed study of this subject goes beyond the scope of
this paper. However, it is demonstrated above on the example of $\hat{D}_4$, that the procedure of
constructing the Poisson submanifolds in affine Lie groups proposed in \cite{AMJGP,FM} works
beyond the case of $\hat{A}$-series as well. We have also constructed the spectral curves for all
$\hat{D}_N$ relativistic Toda chains, using the co-extended affine Lie group generators, their Newton
polygons form a family, presented at fig.~\ref{fi:NpolygonDhat}. The
corresponding integrals of motion for the particular limit of the Casimir functions turn into
corresponding formulas of the non-affine case. We hope to return to these issues elsewhere.

\section*{Acknowledgements}

We are deeply indebted to V.~Fock and E.~Feigin for illuminating discussions. These results have been partially presented at \emph{4th Workshop on Combinatorics of moduli spaces and cluster algebras} in Moscow, May 2014, workshop
\emph{Integrability and Isomonodromy in Mathematical Physics}, at the Lorentz center in Leiden, July 2014,
conference
\emph{Integrability and Cluster Algebras: Geometry and Combinatorics}, ICERM, Brown University in
Providence RI, August 2014, and \emph{Summer School on Cluster Algebras in Mathematical Physics}, Graduate School of Mathematics, Nagoya University, September 2014. We would like to thank their organizers
and participants - in particular M.~Gekhtman, V.~Ovsienko, N.~Reshetikhin, M.~Shapiro, H.~Williams,  M.~Yamazaki and P.~Zinn-Justin for the clarification of some important points, discussed above, and discussion of many related issues.

The work of O.K. was supported by RFBR grant 13-02-00457, the work of A.M. was supported by the
RFBR grant 14-01-00547, by the joint Ukrainian-Russian RFBR project 14-01-90404, and by the Program of Support of Scientific Schools (NSh-1500.2014.2).

\newpage

\appendix

\section*{Appendix}
\def\theequation{\thesection.\arabic{equation}}
\setcounter{equation}0
\section{Some notations
\label{ap:promunu}}

Recall that given a Cartan matrix $C_{ij}$, the Lie algebra $\mathfrak{g}$ is generated by $\{h_i|i \in \Pi\}$ and $\{e_i|i \in \Pi\cup{\bar\Pi}\}$ labeled by simple positive $\Pi$ and negative $\bar{\Pi}$ roots~\footnote{To simplify them we extend $h$ and $C$ to negative values of indices, assuming that $h_i = h_{-i}$ and that $C_{-i,-j}=C_{ij}$ and $C_{ij}=0$ if $i$ and $j$ have different signs, also using the notation $e_{-i}=e_{\bar i}$ for $i>0$.}, subject to the relations
\be\label{rela}
[h_i,h_j]=0,\ \ \ [h_i,e_{j}] = \sign(j)C_{ij} e_j,\ \ \ [e_i,e_{-i}] =\sign(i) h_i,
\\
(\Ad\ e_i)^{1-C_{ij}}e_j = 0 \mbox{ for } i+j\neq 0
  \ee
One can replace the set $\{ h_i\}$ by the dual set $\{h^i\}$ by $h_i=\sum_{j\in\Pi}C_{ij}h^j$, so
that
\be\label{relx}
[h^i,h^j]=0,\ \ \ [h^i,e_{j}] = \sign(j)\delta_i^j e_j,\ \ \ [e_i,e_{-i}] =\sign(i) C_{ij}h^j,
\\
(\Ad\ e_i)^{1-C_{ij}}e_j = 0 \mbox{ for } i+j\neq 0
  \ee
For any $i \in \Pi\cup{\bar\Pi}$ one can introduce the group element $E_i = \exp(e_i)$ and a one-parameter subgroup $H_i(z)=\exp(\log z\cdot h^{i})$. It is important for introducing the cluster coordinates on Poisson submanifolds in Lie groups to use the dual Cartan generators \rf{relx}, this ensures the consistency of the whole construction.

For the $\mathfrak{g}=sl_N$ Lie algebras the Cartan matrix is just
\be
\label{CslN}
C_{ij}\ \stackreb{sl_N}{=}\ (\alpha_i,\alpha_j)= 2\delta_{ij} - \delta_{i+1,j}-\delta_{i,j+1}
\ee
for the positive simple roots $\alpha_i\in\Pi$, $i,j=1,\ldots,\rank\ G=N-1$. Generally
\be
\label{Cgen}
C_{ij} = \langle\alpha_i,\alpha_j\rangle= {2\over (\alpha_i,\alpha_i)}(\alpha_i,\alpha_j)
= {1\over d_i}(\alpha_i,\alpha_j)
\ee
and it is not symmetric for the non-simply-laced case, when roots have different lengths so that
$d_i = \half(\alpha_i,\alpha_i)$ are not necessarily unities. Instead
of the Cartan matrix, which itself defines the Poisson structure on Toda symplectic leaves in the
simply-laced cases, for the non-simply-laced groups one should use instead its symmetrization (see e.g. \cite{Kac}), defined as
\be
C_{ij} = d_iB_{ij},\ \ \ \ B_{ij} = {1\over d_id_j}(\alpha_i,\alpha_j),\ \ \ \ i,j\in\Pi
\ee
The dual vectors
\be
\label{amu}
\langle\alpha_i,\mu_j\rangle=\delta_{ij}
\ee
are the highest weights of the fundamental representations $\pi_{\mu_j}$.

\setcounter{equation}0
\section{Explicit formulas for the integrals of motion
\label{ap:explicit}}

Here we present a list of somewhat more complicated expressions, too long for the main text.

\subsection{Simply-laced case.}

\paragraph{$D_4$.} The explicit form of the triality-invariant $R_2 = \mathcal{H}_2 = \Tr_{\bmu_2}g({\bf x},{\bf y})$ reads
\be
\label{R2D4}
\mathcal{H}_2=
\frac{1}{y_1x_1y_2^2x_2^2y_3x_3y_4x_4}(y_2^4x_2^4y_1^2x_1^2y_3^2x_3^2y_4^2x_4^2+y_2^4x_2^3y_1^2x_1^2y_3^2x_3^2y_4^2x_4^2+y_2^3x_2^3y_1^2x_1^2y_3^2x_3^2y_4^2x_4^2+
\\
+y_2^3x_2^3y_1^2x_1^2y_3^2x_3^2y_4^2x_4+y_2^3x_2^3y_1^2x_1^2y_3^2x_3y_4^2x_4^2+y_2^3x_2^3y_1^2x_1y_3^2x_3^2y_4^2x_4^2+y_2^3x_2^3y_1^2x_1^2y_3^2x_3^2y_4x_4+
\\
+y_2^3x_2^3y_1^2x_1^2y_3^2x_3y_4^2x_4+y_2^3x_2^3y_1^2x_1^2y_3x_3y_4^2x_4^2+y_2^3x_2^3y_1^2x_1y_3^2x_3^2y_4^2x_4+y_2^3x_2^3y_1^2x_1y_3^2x_3y_4^2x_4^2+
\\
y_2^3x_2^3y_1x_1y_3^2x_3^2y_4^2x_4^2+y_2^3x_2^3y_1^2x_1^2y_3^2x_3y_4x_4+y_2^3x_2^3y_1^2x_1^2y_3x_3y_4^2x_4+y_2^3x_2^3y_1^2x_1y_3^2x_3^2y_4x_4+
\\
y_2^3x_2^3y_1^2x_1y_3^2x_3y_4^2x_4+y_2^3x_2^3y_1^2x_1y_3x_3y_4^2x_4^2+y_2^3x_2^3y_1x_1y_3^2x_3^2y_4^2x_4+y_2^3x_2^3y_1x_1y_3^2x_3y_4^2x_4^2+
\\
+y_2^3x_2^3y_1^2x_1^2y_3x_3y_4x_4+y_2^3x_2^3y_1^2x_1y_3^2x_3y_4x_4+y_2^3x_2^3y_1^2x_1y_3x_3y_4^2x_4+y_2^3x_2^3y_1x_1y_3^2x_3^2y_4x_4+
\\
+y_2^3x_2^3y_1x_1y_3^2x_3y_4^2x_4+y_2^3x_2^3y_1x_1y_3x_3y_4^2x_4^2+y_2^3x_2^3y_1^2x_1y_3x_3y_4x_4+y_2^3x_2^3y_1x_1y_3^2x_3y_4x_4+
\\
+y_2^3x_2^3y_1x_1y_3x_3y_4^2x_4+y_2^3x_2^2y_1^2x_1^2y_3x_3y_4x_4+y_2^3x_2^2y_1x_1y_3^2x_3^2y_4x_4+y_2^3x_2^2y_1x_1y_3x_3y_4^2x_4^2+
\\
+y_2^3x_2^3y_1x_1y_3x_3y_4x_4+y_2^3x_2^2y_1^2x_1y_3x_3y_4x_4+y_2^3x_2^2y_1x_1y_3^2x_3y_4x_4+y_2^3x_2^2y_1x_1y_3x_3y_4^2x_4+
\\
+y_2^2x_2^2y_1^2x_1^2y_3x_3y_4x_4+y_2^2x_2^2y_1x_1y_3^2x_3^2y_4x_4+y_2^2x_2^2y_1x_1y_3x_3y_4^2x_4^2+2y_2^3x_2^2y_1x_1y_3x_3y_4x_4+
\\
+2y_2^2x_2^2y_1^2x_1y_3x_3y_4x_4+2y_2^2x_2^2y_1x_1y_3^2x_3y_4x_4+2y_2^2x_2^2y_1x_1y_3x_3y_4^2x_4+y_2^3x_2y_1x_1y_3x_3y_4x_4+
\\
+y_2^2x_2^2y_1^2y_3x_3y_4x_4+y_2^2x_2^2y_1x_1y_3^2y_4x_4+y_2^2x_2^2y_1x_1y_3x_3y_4^2+4y_2^2x_2^2y_1x_1y_3x_3y_4x_4+
\\
+2y_2^2x_2^2y_1x_1y_3x_3y_4+2y_2^2x_2^2y_1x_1y_3y_4x_4+2y_2^2x_2^2y_1y_3x_3y_4x_4+2y_2^2x_2y_1x_1y_3x_3y_4x_4+
\\
+y_2^2x_2^2y_1x_1y_3x_3+y_2^2x_2^2y_1x_1y_4x_4+y_2^2x_2^2y_3x_3y_4x_4+y_2^2x_2y_1x_1y_3x_3y_4+
\\
+y_2^2x_2y_1x_1y_3y_4x_4+y_2^2x_2y_1y_3x_3y_4x_4+y_2x_2y_1x_1y_3x_3y_4x_4+y_2^2x_2y_1x_1y_3x_3+
\\
+y_2^2x_2y_1x_1y_4x_4+y_2^2x_2y_3x_3y_4x_4+y_2x_2y_1x_1y_3x_3y_4+y_2x_2y_1x_1y_3y_4x_4+y_2x_2y_1y_3x_3y_4x_4+
\\
+y_2x_2y_1x_1y_3x_3+y_2x_2y_1x_1y_3y_4+y_2x_2y_1x_1y_4x_4+y_2x_2y_1y_3x_3y_4+y_2x_2y_1y_3y_4x_4+
\\
+y_2x_2y_3x_3y_4x_4+y_2x_2y_1x_1y_3+y_2x_2y_1x_1y_4+y_2x_2y_1y_3x_3+y_2x_2y_1y_3y_4+y_2x_2y_1y_4x_4+
\\
+y_2x_2y_3x_3y_4+y_2x_2y_3y_4x_4+y_2x_2y_1x_1+y_2x_2y_1y_3+y_2x_2y_1y_4+
\\
+y_2x_2y_3x_3+y_2x_2y_3y_4+y_2x_2y_4x_4+y_2x_2y_1+y_2x_2y_3+y_2x_2y_4+y_2x_2+y_2+1)
\ee

\paragraph{$E_6$.} The explicit form of $\mathcal{H}_2 = \Tr_{\bmu_2}g({\bf x},{\bf y})$,
corresponding to the trace in 78-dimensional adjoint representation is
\be
\mathcal{H}_2=\frac{1}{y_6x_6 y_5^2x_5^2 y_4^3x_4^3 y_3^2x_3^2 y_2^2x_2^2 y_1x_1}\cdot\mathcal{P}({\bf x},{\bf y})
\ee
where $\mathcal P$ is some polynomial of $y_i,\  x_i, \ i=1,..,6$. The monomial factor in this expression is encoded by the matrix elements $C_{2j}$ of $E_6$.

\subsection{Non simply-laced}

\paragraph{$B_3$.} This integral of motion is obtained from the formula \rf{R2D4}
by reduction $x_3=x_4$ and $y_3=y_4$
\be
\label{R2B3}
\mathcal{H}_2=\frac{1}{y_3^2x_3^2y_2^2x_2^2y_1x_1}(y_2^4x_2^4y_1^2x_1^2y_3^4x_3^4+y_2^4x_2^3y_1^2x_1^2y_3^4x_3^4+y_2^3x_2^3y_1^2x_1^2y_3^4x_3^4+2y_2^3x_2^3y_1^2x_1^2y_3^4x_3^3+
\\
+y_2^3x_2^3y_1^2x_1y_3^4x_3^4+y_2^3x_2^3y_1^2x_1^2y_3^4x_3^2+2y_2^3x_2^3y_1^2x_1^2y_3^3x_3^3+
2y_2^3x_2^3y_1^2x_1y_3^4x_3^3+y_2^3x_2^3y_1x_1y_3^4x_3^4+
\\
+2y_2^3x_2^3y_1^2x_1^2y_3^3x_3^2+y_2^3x_2^3y_1^2x_1y_3^4x_3^2+2y_2^3x_2^3y_1^2x_1y_3^3x_3^3+2y_2^3x_2^3y_1x_1y_3^4x_3^3+
y_2^3x_2^3y_1^2x_1^2y_3^2x_3^2+
\\
+y_2^3x_2^2y_1^2x_1^2y_3^2x_3^2+2y_2^3x_2^2y_1x_1y_3^3x_3^3+y_2^3x_2^3y_1x_1y_3^2x_3^2+y_2^3x_2^2y_1^2x_1y_3^2x_3^2+2y_2^3x_2^2y_1x_1y_3^3x_3^2+
\\
+y_2^2x_2^2y_1^2x_1^2y_3^2x_3^2+
2y_2^2x_2^2y_1x_1y_3^3x_3^3+2y_2^3x_2^2y_1x_1y_3^2x_3^2+
2y_2^2x_2^2y_1^2x_1y_3^2x_3^2+
\\
+4y_2^2x_2^2y_1x_1y_3^3x_3^2+y_2^3x_2y_1x_1y_3^2x_3^2+y_2^2x_2^2y_1^2y_3^2x_3^2+
2y_2^2x_2^2y_1x_1y_3^3x_3+4y_2^2x_2^2y_1x_1y_3^2x_3^2+
\\
+4y_2^2x_2^2y_1x_1y_3^2x_3+2y_2^2x_2^2y_1y_3^2x_3^2+2y_2^2x_2y_1x_1y_3^2x_3^2+2y_2^2x_2^2y_1x_1y_3x_3+y_2^2x_2^2y_3^2x_3^2+
\\
+2y_2^2x_2y_1x_1y_3^2x_3+y_2^2x_2y_1y_3^2x_3^2+y_2x_2y_1x_1y_3^2x_3^2+2y_2^2x_2y_1x_1y_3x_3+
y_2^2x_2y_3^2x_3^2+
\\
+2y_2x_2y_1x_1y_3^2x_3+y_2x_2y_1y_3^2x_3^2+y_2x_2y_1x_1y_3^2+
2y_2x_2y_1x_1y_3x_3+
2y_2x_2y_1y_3^2x_3+
\\
+y_2x_2y_3^2x_3^2+2y_2x_2y_1x_1y_3+y_2x_2y_1y_3^2+2y_2x_2y_1y_3x_3+2y_2x_2y_3^2x_3+
y_2x_2y_1x_1+
\\
+2y_2x_2y_1y_3+y_2x_2y_3^2+2y_2x_2y_3x_3+y_2x_2y_1+2y_2x_2y_3+y_2x_2+y_2+1)
\ee
\paragraph{ $C_3$.}
Two extra integrals of motion in this case - in addition to \rf{C3first} are
\be
\label{R23C3}
\mathcal{H}_2=\mathcal{H}_4=\frac{1}{y_1y_2^2y_3x_1x_2^2x_3} (y_1^2 y_2^4 y_3^2 x_1^2 x_2^4 x_3^2+y_1^2 y_2^4 y_3^2 x_1^2 x_2^3 x_3^2+y_1^2 y_2^3 y_3^2 x_1^2 x_2^3 x_3^2+y_1^2 y_2^3 y_3^2 x_1^2 x_2^3 x_3+
\\
y_1^2 y_2^3 y_3^2 x_1 x_2^3 x_3^2+y_1^2 y_2^3 y_3^2 x_1 x_2^3 x_3+y_1^2 y_2^3 y_3 x_1^2 x_2^3 x_3+y_1 y_2^3 y_3^2 x_1 x_2^3 x_3^2+y_1^2 y_2^3 y_3 x_1^2 x_2^2 x_3+
\\
y_1^2 y_2^3 y_3 x_1 x_2^3 x_3+y_1 y_2^3 y_3^2 x_1 x_2^3 x_3+
y_1^2 y_2^3 y_3 x_1 x_2^2 x_3+y_1^2 y_2^2 y_3 x_1^2 x_2^2 x_3+y_1 y_2^3 y_3 x_1 x_2^3 x_3+
\\
2 y_1^2 y_2^2 y_3 x_1 x_2^2 x_3+
2 y_1 y_2^3 y_3 x_1 x_2^2 x_3+y_1^2 y_2^2 y_3 x_2^2 x_3+y_1 y_2^3 y_3 x_1 x_2 x_3+3 y_1 y_2^2 y_3 x_1 x_2^2 x_3+
\\
2 y_1 y_2^2 y_3 x_1 x_2 x_3+2 y_1 y_2^2 y_3 x_2^2 x_3+y_1 y_2^2 y_3 x_2 x_3+y_1 y_2 y_3 x_1 x_2 x_3+y_2^2y_3x_2^2x_3+y_1y_2y_3x_1x_2+
\\
y_1y_2y_3x_2x_3+y_2^2y_3x_2x_3+y_1y_2y_3x_2+y_1y_2x_1x_2+y_2y_3x_2x_3+y_1y_2x_2+y_2y_3x_2+y_2x_2+y_2+1)
\ee
and
\be
\mathcal{H}_3=\frac{1}{y_1y_2^2y_3^{3/2}x_1x_2^2x_3^{3/2}}  (  y_1^2 y_2^4 y_3^3 x_1^2 x_2^4 x_3^3+y_1^2 y_2^4 y_3^3 x_1^2 x_2^4 x_3^2+
y_1^2 y_2^4 y_3^2 x_1^2 x_2^4 x_3^2+2 y_1^2 y_2^4 y_3^2 x_1^2 x_2^3 x_3^2+
\\
y_1^2 y_2^4 y_3^2 x_1^2 x_2^2 x_3^2+2 y_1^2 y_2^3 y_3^2 x_1^2 x_2^3 x_3^2+2 y_1^2 y_2^3 y_3^2 x_1^2 x_2^2 x_3^2+
2 y_1^2 y_2^3 y_3^2 x_1 x_2^3 x_3^2+2 y_1^2 y_2^3 y_3^2 x_1 x_2^2 x_3^2+
\\
y_1^2 y_2^2 y_3^2 x_1^2 x_2^2 x_3^2+2 y_1 y_2^3 y_3^2 x_1 x_2^3 x_3^2+y_1^2 y_2^2 y_3^2 x_1^2 x_2^2 x_3+2 y_1^2 y_2^2 y_3^2 x_1 x_2^2 x_3^2+2 y_1 y_2^3 y_3^2 x_1 x_2^2 x_3^2+
\\
2 y_1^2 y_2^2 y_3^2 x_1 x_2^2 x_3+y_1^2 y_2^2 y_3^2 x_2^2 x_3^2+y_1^2 y_2^2 y_3 x_1^2 x_2^2 x_3+2 y_1 y_2^2 y_3^2 x_1 x_2^2 x_3^2+y_1^2 y_2^2 y_3^2 x_2^2 x_3+
\\
2 y_1^2 y_2^2 y_3 x_1 x_2^2 x_3+2 y_1 y_2^2 y_3^2 x_1 x_2^2 x_3+2 y_1 y_2^2 y_3^2 x_2^2 x_3^2+
y_1^2 y_2^2 y_3 x_2^2 x_3+2 y_1 y_2^2 y_3^2 x_2^2 x_3+
\\
2 y_1 y_2^2 y_3 x_1 x_2^2 x_3+y_2^2 y_3^2 x_2^2 x_3^2+2 y_1 y_2^2 y_3 x_1 x_2 x_3+
2 y_1 y_2^2 y_3 x_2^2 x_3+y_2^2 y_3^2 x_2^2 x_3+
\\
2 y_1 y_2^2 y_3 x_2 x_3+2y_1y_2y_3x_1x_2x_3+
y_2^2y_3x_2^2x_3+2y_1y_2y_3x_2x_3+
\\
2y_2^2y_3x_2x_3+y_2^2y_3x_3+2y_2y_3x_2x_3+2y_2y_3x_3+y_3x_3+y_3+1  )
\ee


\newpage


\begin{thebibliography}{9}

\bibitem{Sembook} A.~Reiman and M.~Semenov-Tian-Shansky, \textit{Integrable systems}, 2003.

\bibitem{FMold}
V.~V.~Fock and A.~Marshakov, {\em A Note on Quantum Groups and Relativistic Toda Theory},
Nucl.Phys. {\bf 56B} (Proc. Suppl.) (1997) 208-214.

\bibitem{AMJGP}
A.~Marshakov, \emph{Lie Groups, Cluster Variables and Integrable Systems}, J. Geom. Phys. 67 (2013) 16–36, \texttt{arXiv:1207.1869 [hep-th]}.

\bibitem{FM}
V.~V.~Fock and A.~Marshakov, {\em Loop Groups, Clusters, Dimers and Integrable Systems}, \texttt{arXiv:1401.1606}

\bibitem{GK}
A.~Goncharov and R.~Kenyon,
    \textit{Dimers and cluster integrable systems}, \texttt{arXiv:1107.5588}.


\bibitem{cv}
S.~Fomin and A.~Zelevinsky,  \emph{Cluster algebras I: Foundations}, \texttt{arXiv:math/0104151};
\\
A.~Berenstein, S.~Fomin and A.~Zelevinsky, \emph{Cluster algebras III: Upper bounds and double Bruhat cells}, \texttt{arXiv:math/0305434};
    \\
V.~V.~Fock and A.~B.~Goncharov, \emph{Moduli spaces of local systems and higher Teichmuller theory},
 \texttt{arXiv:math/0311149}.


\bibitem{bruh}
T.~Hodges and T.~Levasseur, 
Comm.~Math.~Phys. {\bf 156} (1993) 561;\\
T.~Hoffman, J.~Kellendonk, N.~Kuntz and N.~Reshetikhin,
Comm.~Math.~Phys.  \textbf{212}  (2000) 297, \texttt{arXiv:solv-int/9906013};\\
M.~Kogan, A.~Zelevinsky,
Int. Math. Res. Not. 2002, no. 32, 1685--1702, \texttt{arXiv:math/0203069}.

\bibitem{Resh}
N.~Reshetikhin, \emph{Characteristic systems on Poisson Lie groups and their quantization}, in \emph{Integrable systems: from classical to quantum} (Montre'al, QC, 1999), 165–188, CRM Proc. Lecture Notes, 26, Amer. Math. Soc., Providence, RI, 2000; arXiv:0103147.


\bibitem{GSW}
M.~Gekhtman, M.~Shapiro and A.~Vainshtein,
\emph{Exotic cluster structures on SL(n): the Cremmer-Gervais case}, arXiv:1307.1020;
\\
M.~Gekhtman, M.~Shapiro and A.~Vainshtein,
\emph{Cremmer-Gervais cluster structure on SL(n)},  arXiv:1308.2558.

\bibitem{GSWbook}
M.~Gekhtman, M.~Shapiro and A.~Vainshtein,
\emph{Cluster Algebra and Poisson Geometry}, AMS-2010.

\bibitem{HW}
H.~Williams,
\emph{Double Bruhat Cells in Kac-Moody Groups and Integrable Systems},  arXiv:1204.0601;
\\
H.~Williams,
\emph{Cluster Ensembles and Kac-Moody Groups}, Adv. Math. \textbf{247} (2013), pp. 1-40,
 arXiv:1210.2533;
\\
H.~Williams,
\emph{Q-Systems, Factorization Dynamics, and the Twist Automorphism},  arXiv:1310.6624.

\bibitem{pent}
R. Schwartz, \textit{The pentagram map is recurrent,} Experiment. Math. 10 (2001), 519--528;
\\
V.~Ovsienko, R.~Schwartz and S.~Tabachnikov,
\textit{The Pentagram map: a discrete integrable system},  Comm. Math. Phys.  299  (2010),  no. 2, 409--446, \texttt{arXiv:0810.5605};
\\
V.~Ovsienko, R.~Schwartz and S.~Tabachnikov,
\textit{Liouville-Arnold integrability of the pentagram map on closed polygons}, \texttt{arXiv:1107.3633.};
\\
M.Glick, \textit{The pentagram map and Y-patterns},  Adv.Math.  227  (2011),  no. 2, 1019-1045, \texttt{arXiv:1005.0598};
\\
M.~Gekhtman, M.~Shapiro, S.~Tabachnikov and A.~Vainshtein,
\textit{ Higher pentagram maps, weighted directed networks, and cluster dynamics},  Electron. Res. Announc. Math. Sci.  19  (2012), 1--17. \texttt{arXiv:1110.0472}.

\bibitem{SolKhe}
F.~Soloviev, \textit{Integrability of the Pentagram Map}, \texttt{arXiv:1106.3950};
\\
B.~Khesin, F.~Soloviev, \textit{The pentagram map in higher dimensions and KdV flows.} Electron. Res. Announc. Math. Sci.  19  (2012), 86--96, \texttt{arXiv:1205.3744}, \textit{Integrability of higher pentagram maps,} \texttt{arXiv:1204.0756}.


\bibitem{HWN}
H.~Williams, in \emph{Toda systems, quiver representations, and N=2 field theory},
Lectures at Summer school on cluster algebras in mathematical physics, Nagoya 2014.

\bibitem{Ruj}
S.Ruijsenaars, Comm.Math.Phys. {\bf 133} (1990) 217.

\bibitem{Suris} Yu.B.~Suris, \textit{Algebraic structure of discrete-time and relativistic Toda lattices,} Phys.Lett. A156 (1991), 467-474.



\bibitem{DrinFG} V.~Fock and A.~Goncharov, {\it Cluster X-varieties, amalgamation and Poisson-Lie groups}. In {\it Algebraic Geometry Theory and Number Theory}, pp. 27--68, Progr. Math., 253, Birkh\"auser Boston, Boston, MA, 2006. {\tt math.RT/0508408}.

\bibitem{vinb}
E.~B.~Vinberg, A.~L.~Onischik, \emph{Seminar on Lie groups and algebraic groups}, Nauka, Moscow, 1988

\bibitem{prog}
Lie online service:
http://www-math.univ-poitiers.fr/~maavl/LiE/form.html

\bibitem{Kac}
V.~Kac, \emph{Infinite-dimensional Lie algebras}, Cambridge University Press




\end{thebibliography}
\end{document}